\documentclass[acmsmall]{acmart}

\setcopyright{cc}
\setcctype{by}
\acmDOI{10.1145/3763163}
\acmYear{2025}
\acmJournal{PACMPL}
\acmVolume{9}
\acmNumber{OOPSLA2}
\acmArticle{385}
\acmMonth{10}
\received{2025-03-26}
\received[accepted]{2025-08-12}

\usepackage[utf8]{inputenc}
\usepackage{enumitem}
\usepackage{xspace}
\usepackage[linesnumbered,vlined]{algorithm2e}
\usepackage[table]{xcolor}
\usepackage{wrapfig}
\usepackage{listings}
\usepackage{multirow}
\usepackage{pifont}
\usepackage{setspace}
\AtBeginDocument{%
  }

\newcommand{\ignore}[1]{}

\newcommand{\ie}{i.e.,\xspace}
\newcommand{\aka}{a.k.a.\xspace}
\newcommand{\eg}{e.g.,\xspace}
\newcommand{\triple}[3]{\{#1\}{\scalebox{0.9}{$#2$}}\{#3\}}
\newcommand{\tool}{\textsc{AutoBug}\xspace}
\newcommand{\unfold}{\mathit{unfold}}

\newcommand{\pp}{\pi}
\newcommand{\PP}{\mathcal{P}}
\newcommand{\QQ}{\mathcal{Q}}

\newcommand{\spc}{\mathit{sp}}
\newcommand{\skipc}{\mathbf{skip}}
\newcommand{\itec}[3]{\mathbf{if}~#1~\mathbf{then}~#2~\mathbf{else}~#3~\mathbf{end}}
\newcommand{\whilec}[2]{\mathbf{while}~#1~\mathbf{do}~#2~\mathbf{done}}
\newcommand{\assertc}[1]{\mathbf{assume}(#1)}
\newcommand{\assumec}[1]{\mathbf{assume}(#1)}

\newcommand{\vars}[1]{\mathit{vars}(#1)}
\newcommand{\cov}[1]{\mathit{cov}(#1)}
\newcommand{\myparagraph}[1]{\vspace{0.5em}\noindent\emph{#1.}}
\newcommand{\rn}[1]{\expandafter{\romannumeral #1\relax}}

\newcommand{\cmark}[0]{\color{green!50!black}{\ding{51}}}
\newcommand{\xmark}[0]{\color{red!50!black}{\ding{55}}}
\newcommand{\rev}[1]{#1}

\theoremstyle{definition}

\newtheorem{example}{Example}[section]

\renewcommand{\equationautorefname}{\relax}
\def\equationautorefname#1{}

\definecolor{mycolor}{rgb}{0.122, 0.435, 0.698}
\usepackage{tcolorbox}
\newcommand{\result}[1]{%
\begin{tcolorbox}[colframe=mycolor,boxrule=0.5pt,arc=4pt,
      left=6pt,right=6pt,top=6pt,bottom=6pt,boxsep=0pt,width=\columnwidth]%
      {#1}
\end{tcolorbox}%
}

\begin{document}

\title{Large Language Model Powered Symbolic Execution}

\author{Yihe Li}
\orcid{0009-0008-2257-0406}
\affiliation{%
  \institution{National University of Singapore}
  \city{Singapore}
  \country{Singapore}
}
\email{yihe.li@u.nus.edu}

\author{Ruijie Meng}
\orcid{0009-0008-4323-6271}
\authornote{Corresponding author}
\affiliation{%
  \institution{National University of Singapore}
  \city{Singapore}
  \country{Singapore}
}
\email{ruijie@comp.nus.edu.sg}

\author{Gregory J. Duck}
\orcid{0000-0002-0837-9671}
\affiliation{%
  \institution{National University of Singapore}
  \city{Singapore}
  \country{Singapore}
}
\email{gregory@comp.nus.edu.sg}




\begin{abstract}
{\em Large Language Models} (LLMs) have emerged as a promising alternative to traditional static program analysis methods, such as symbolic execution, offering the ability to reason over code {\em directly} without relying on theorem provers or SMT solvers.
However, LLMs are also inherently approximate by nature, and therefore face significant challenges in relation to the {\em accuracy} and {\em scale} of analysis in real-world applications.
Such issues often necessitate the use of larger LLMs with higher token limits, but this requires enterprise-grade hardware (GPUs) and thus limits accessibility for many users. In this paper, we propose {\em LLM-based symbolic execution}---a novel approach that enhances LLM inference via a path-based decomposition of the program analysis tasks into smaller (more tractable) subtasks.
The core idea is to generalize path constraints using a generic code-based representation that the LLM can directly reason over, and without translation into another (less-expressive) formal language.
We implement our approach in the form of \tool, an LLM-based symbolic execution engine that is lightweight and language-agnostic, making it a practical tool for analyzing code that is challenging for traditional approaches.
We show that \tool can improve both the accuracy and scale of LLM-based program analysis, especially for smaller LLMs that can run on consumer-grade hardware.
\end{abstract}

\maketitle

\section{Introduction}

{\em Program analysis} is a foundational discipline in computer science that aims to understand program behavior through various systematic techniques.
Traditional forms of program analysis include {\em static} methods---such as {\em symbolic execution}~\cite{King76}, {\em abstract interpretation}~\cite{Cousot77, Cousot79}, and {\em model checking}~\cite{Clarke81}---that analyze the program without executing it, as well as {\em dynamic} methods---such as {\em fuzzing}~\cite{Miller90, fuzzingbook2024}, {\em concolic execution}~\cite{Dart05, Cute05}, {\em instrumentation}~\cite{Asan12, Valgrind07}, and {\em profiling}~\cite{Gprof82}---that analyze program behavior based on observations of actual executions.
In general, program analysis has many applications, such as program testing~\cite{Hamlet77, Clarke76, Rapps85, Zhu97, Dart05, Cadar08, Cute05, Pacheco07, Weyuker83, Goodenough75}, debugging~\cite{Zeller05, Agrawal93, Dawson01}, verification~\cite{Floyd67, Hoare69}, repair~\cite{Genprog, Long16}, reverse engineering~\cite{nelson96}, and vulnerability detection~\cite{Kaur20}.
A recent alternative to traditional program analysis methods has emerged in the form of {\em Large Language Models} (LLMs)~\cite{Chen21, Li22, Llama23, Gpt4, Openai24}.
Here, LLMs can make inferences over code directly with properties expressed in code or natural language and have become powerful enough to handle many traditional program analysis applications.
As such, many LLM-based testing~\cite{Chen24chat, chatafl}, debugging~\cite{Yacine24}, and repair~\cite{Lo24detect, Zhou24, Xia23} tools have recently emerged.

We contrast traditional program analysis (symbolic execution) with LLM-based program analysis.
Here, symbolic execution is a static program analysis method based on the idea of executing a program with {\em symbolic inputs/values}, typically represented as logical formulae over some underlying theories (\eg linear arithmetic, bit vectors, and arrays).
Symbolic execution systematically transforms these symbolic values (\aka the {\em symbolic state}) based on the program's statements, effectively executing multiple {\em concrete} paths simultaneously.
These symbolic states can then be used for various analysis tasks, such as verifying the truth of an assertion or generating test cases, using {\em deductive inference} with the help of some underlying {\em theorem prover} or {\em SMT solver}~\cite{DeMoura08}.
In contrast, LLM-based program analysis involves engineering a {\em prompt} (or sequence of {\em prompts}) that queries an LLM as an oracle.
A typical prompt consists of the relevant code (or code fragments), as well as instructions explaining the analysis task expressed in natural language.
The LLM effectively uses a form of {\em approximate inference} (based on the training data) to solve the task, rather than the strict deductive inference used by a theorem prover.
As such, LLM-based program analysis is typically ad hoc and tailored to specific tasks rather than general and principled.

Both traditional and LLM-based program analysis face significant challenges in practice.
For example, traditional symbolic execution has several well-known limitations~\cite{Baldoni18, Cadar13}, such as the handling of {\em unbounded} loops, the handling of external {\em environment}/libraries, and the handling of memory/{\em heap}-manipulating programs---all of which are common-place in real-world code.
We consider how KLEE~\cite{Cadar08}, a prominent symbolic execution engine, treats each iteration of a loop as a separate path, leading to non-termination for unbounded loops (\eg $\texttt{while(}i~\texttt{<}~\mathit{input}\texttt{)}~\texttt{\{}...\texttt{\}}$). 
In contrast, LLMs have the ability to directly reason over loops, environment, and heaps, avoiding problems such as non-termination.
That said, LLMs face other challenges, such as {\em scalability} (\eg the 8192 {\em token limit} for GPT-4~\cite{Gpt4}) and {\em accuracy}~\cite{Levy24, Fang24} due to approximate reasoning.
Recent studies also correlate the accuracy of LLMs with prompt size~\cite{Levy24}, meaning that more concise and targeted prompts generally perform better.

In this paper, our aim is to improve the accuracy and scale of LLM-based program analysis.
Our first main insight is that the strengths/weaknesses of traditional and LLM-based program analysis are {\em complementary}, meaning that a hybrid design can help address the limitations of either approach.
To this end, we propose combining the path-based decomposition of symbolic execution with approximate inference via an LLM---\aka~{\em LLM-based symbolic execution}.
The core idea is a principled decomposition of the original program analysis task into smaller subtasks based on paths in the original program---helping to mitigate some of the scalability and accuracy concerns with LLMs.
Our second main insight is that since LLMs are primarily trained on code, the representation of symbolic states should also be in terms of code rather than logical formulae.
To this end, we propose a generic path constraint representation of the form of a {\em strongest post-condition} ($\spc$) predicate transformer~\cite{Dijkstra75} over {\em sub-programs} derived from the original program, where each sub-program represents a path.
Since we represent path constraints as ordinary code, we can use an LLM prompting {\em directly}, disposing of any {\em verification conditions} (VCs) generated by the symbolic execution process.
Our approach also avoids many of the limitations associated with the translation of paths into formulae for a theorem prover (\eg environment and heaps).
Our final insight is that our path constraint representation can be {\em generalized} into {\em sets} of paths (\eg all iterations of an unbounded loop), ensuring that LLM-based symbolic execution will always terminate.

We study LLM-based symbolic execution in both theory and practice.
We show that LLM-based symbolic execution mitigates many of the limitations of both traditional and LLM-based program analysis.
We have implemented our approach in the form of \tool---an automated LLM-based symbolic execution engine.
\tool uses a path-based decomposition of a program analysis task into smaller (more tractable) subtasks that are suitable for LLM inference.
Based on the observation that LLMs are inherently approximate oracles, we propose a lightweight design of \tool that is language agnostic and does not rely on any heavyweight compiler infrastructure.
\tool is designed to be a practical program analysis tool that can be applied to code that is difficult to analyze with traditional methods.
In summary, the main contributions of this paper are:
\begin{itemize}[leftmargin=*]
\item We introduce the concept of {\em LLM-based symbolic execution}---a symbolic execution methodology using LLMs for {\em direct} reasoning over the original programming language (\eg \verb+C+/Java/Python), rather than indirect reasoning via translation into some formal theorem prover language.
We show that our approach avoids many well-known limitations of traditional symbolic execution, such as unbounded loops, external environment, heap manipulation, etc.
Furthermore, our approach uses a path-based decomposition of program analysis tasks into more concise LLM prompts, improving the accuracy and scalability of LLM inference.

\item We implement our approach in the form of \tool---a lightweight LLM-based symbolic execution engine that supports multiple programming languages without relying on heavyweight compiler infrastructure.

\item We evaluate \tool against various program analysis tasks for \verb+C+/Python/Java code.
We show that \tool improves both {\em accuracy} and {\em scale}, especially for smaller models that can run on consumer-grade hardware.
\end{itemize}
\section{Motivation}

\subsection{Background}

\subsubsection{Symbolic Execution}\label{sec:symexe_intro}

Symbolic execution~\cite{King76, Baldoni18} is an established program analysis methodology based on running the program with {\em symbolic values} representing {\em sets} of concrete inputs (rather than a single concrete input for {\em normal} execution).
For each symbolic input, symbolic execution will systematically explore multiple execution paths of the program, allowing for applications such as error detection, verifying correctness, or finding vulnerabilities.

Symbolic execution works with {\em symbolic states}
that are traditionally represented as a set of {\em variables} (\eg \verb+x+ and \verb+y+) subject to a {\em path constraint} (\eg $\texttt{x} < \texttt{y}$).
The symbolic state represents the set of {\em concrete} states (\eg $\{(\texttt{x}, \texttt{y})~|~\texttt{x} < \texttt{y}\}$) that is reachable from some symbolic input.
Symbolic execution can be defined in terms of operations over symbolic states, \eg symbolic execution over \verb+C+-style increment statement (\verb_x++_) can be represented as the Hoare triple~\cite{Hoare69}:

\vspace{-1em}
{\small
\begin{align*}
\triple{\PP}{\texttt{ x++ }}{\exists z : \texttt{x} = z + 1 \land \PP[\texttt{x} \mapsto z]}
\end{align*}
}

\noindent
Given a pre-condition ($\PP$), the triple describes the resulting post-condition after the execution of the increment operation.
For example, given a path constraint $\pp_\mathit{pre} = (\texttt{x} < \texttt{y})$, then the path constraint after the operation will be
$\pp_\mathit{post} = (\exists z : \texttt{x} = z + 1 \land z < \texttt{y})$, or equivalently, $\pp_\mathit{post} = (\texttt{x} \leq \texttt{y})$.
Conditional statements (\eg $\itec{c}{T}{E}$) are typically handled by {\em forking} the path constraint ($\pi_\mathit{pre}$) into separate $\pi_\mathit{then} = (\pi_\mathit{pre} \land c)$ and $\pi_\mathit{else} = (\pi_\mathit{pre} \land \neg c)$ constraints, and then continuing execution along the individual {\em then}- and {\em else}-branches.
A similar strategy is used for loops (\eg $\whilec{c}{B}$).
Finally, for each path through the program, a final path constraint $\pp$ will be generated.
This can be used to prove properties over the path, such as whether a final {\em post-condition} ($\QQ$) holds, \ie whether the {\em verification condition} (VC) of the form ($\pp \models \QQ$) holds or not.
Such VCs are discharged with the help of {\em theorem provers}, such as SMT solvers~\cite{DeMoura08}.
By executing sets of concrete inputs at once, symbolic execution aims to exhaustively explore the space of program behaviors---something that cannot be achieved by concrete execution alone.

\begin{example}\label{ex:simple}
Consider the simple program ($\itec{\texttt{x} > \texttt{y}}{\texttt{z} := \texttt{x} + 2}{\texttt{z} := \texttt{x} * \texttt{y}}$) and the post-condition $\QQ = (\texttt{z} > \texttt{y})$.
Then the following VC will be generated for the $\mathbf{else}$-path:
\begin{align}
    \neg (\texttt{x} > \texttt{y}) \land \texttt{z} = \texttt{x} \times \texttt{y} \qquad \models \qquad \texttt{z} > \texttt{y}
    \label{eq:vc1}
    \tag{\sc VC-Else}
\end{align}
A theorem prover shows that (\ref{eq:vc1}) is unsatisfiable, meaning that ($\QQ$) does {\bf not} hold for all executions of the program.
$\qed$\end{example}

\subsubsection{Large Language Models}

The emergence of {\em Large Language Models} (LLMs)~\cite{Chen21, Li22, Llama23, Gpt4, Openai24} presents a new alternative for reasoning over program code.
Trained on extensive datasets, including millions of lines of real-world code written in multiple languages, LLMs can reason over \verb|C|/Java/Python/etc. code directly, including many traditional program analysis tasks, such as
testing~\cite{Chen24chat, chatafl}, debugging~\cite{Yacine24}, and repair~\cite{Lo24detect, Zhou24, Xia23}.
For example, instead of relying on symbolic execution, LLMs can reason over code using a suitable prompt expressed in natural language, such as ``{\em there is a bug present in the following Python code segment, please suggest the possible root causes of the bug and corresponding fixes}''.
Most modern LLMs can analyze the code and generate possible suggestions and patches automatically, based on the understanding of code and defects present in the training set.

Unlike traditional program analysis methods, LLMs do not aim to be perfectly {\em precise}.
Rather, LLMs can be thought of as {\em approximate} oracles that are sometimes incomplete or give the wrong answer.
This is because LLMs fundamentally rely on learned patterns and approximate reasoning, rather than classical deductive reasoning used by traditional theorem provers.
Despite the difference, LLMs are clearly useful, with an explosion of applications for real-world analysis problems.

\lstdefinestyle{customc}{
    language=C,
    basicstyle=\ttfamily\tiny,
    keywordstyle=\color{blue}\bfseries,
    commentstyle=\color{gray},
    stringstyle=\color{red},
    numbers=none,
    numberstyle=\tiny\color{gray},
    stepnumber=1,
    breaklines=true,
    frame=single,
    tabsize=4,
    showstringspaces=false    
}

\lstset{style=customc}

\begin{figure}
\centering
\begin{minipage}{0.47\textwidth}
\centering
{\footnotesize \textcolor{red!60!black}{(a)} original program}
\begin{lstlisting}
#define SIZE 100
typedef struct NODE {
    const char *key, *val;
    struct NODE *next;
} NODE;
NODE *db = NULL;
pthread_mutex_t mutx = PTHREAD_MUTEX_INITIALIZER;

void handle_client(int s) {
  char buf[SIZE], rep[SIZE];
  while (true) {
    ssize_t r = recv(s, buf, SIZE-1, 0);
    if (r <= 0) break;
    buf[r] = '\0';
    char cmd[SIZE] = {0}, key[SIZE], val[SIZE];
    sscanf(buf, "%s %s %s", cmd, key, val);    
    if (strcmp(cmd, "GET") == 0) {
      snprintf(rep, SIZE, "ERROR\n");
      NODE *n, *p = NULL;
      pthread_mutex_lock(&mutx);
      for (n = db; n != NULL; p = n, n = n->next) {
        if (strcmp(n->key, key) != 0)
          continue;
        if (p != NULL) p->next = n->next;
        else           db = n->next;
      }
      pthread_mutex_unlock(&mutx);
      if (n != NULL) {
        snprintf(rep, SIZE, "%s\n", n->val);
        free(n->key); free(n->val); free(n);
      } else
        snprintf(rep, SIZE, "ERROR\n");
    } else if (strcmp(cmd, "PUT") == 0) {
      NODE *n = (NODE *)malloc(sizeof(NODE));
      pthread_mutex_lock(&mutx);
      n->key = strdup(key);
      n->val = strdup(val);
      n->next = db; db = n;
      pthread_mutex_unlock(&mutx);
      snprintf(rep, SIZE, "OK\n");
    } else
      snprintf(rep, SIZE, "ERROR\n");
    send(s, rep, strlen(rep), 0);
  }
  close(s);
}
\end{lstlisting}
\end{minipage}
\hspace{0.2cm}
\begin{minipage}{0.45\textwidth}
\centering
{\footnotesize \textcolor{red!60!black}{(b)} slice for ``\texttt{mutx} is unlocked''}
\begin{lstlisting}[mathescape]
pthread_mutex_t mutx = PTHREAD_MUTEX_INITIALIZER;

void handle_client(int s) {
  while (true) {
    if ($...$) break;
    if ($...$) {
      pthread_mutex_lock(&mutx);
      pthread_mutex_unlock(&mutx);
    } else {
      pthread_mutex_lock(&mutx);
      pthread_mutex_unlock(&mutx);
    }
  }
}
\end{lstlisting}

\vspace{0.5cm}
{\footnotesize \textcolor{red!60!black}{(c)} slice for ``$\texttt{db->key} \neq \texttt{NULL}$''}
\begin{lstlisting}[mathescape]
#define SIZE 100
typedef struct NODE {
    const char *key;
} NODE;
NODE *db = NULL;

void handle_client(int s) {
  char buf[SIZE];
  while (true) {
    ssize_t r = recv(s, buf, SIZE-1, 0);
    if ($...$) break;
    buf[r] = '\0';
    char cmd[SIZE] = {0}, key[SIZE], val[SIZE];
    sscanf(buf, "%s %s %s", cmd, key, val);
    if ($...$) {
      NODE *n = (NODE *)malloc(sizeof(NODE));
      n->key = strdup(key);
      db = n;
    }
  }
}
\end{lstlisting}
\end{minipage}
\caption{An example program \textcolor{red!60!black}{(a)} that implements a simple key-value server.
The example program includes an unbounded loop (\texttt{while (true) ...}), interaction with the external environment (\texttt{recv}/\texttt{send}), and heap-manipulating data structures (\texttt{NODE}).
In addition, slices \textcolor{red!60!black}{(b)} and \textcolor{red!60!black}{(c)} corresponds to the post-conditions ``\texttt{mutx} is unlocked'' (always holds) and ``$\texttt{db->key} \neq \texttt{NULL}$'' (may not hold) respectively.
\label{fig:motivate}}
\end{figure}

\subsection{Limitations of Symbolic Execution}\label{sec:se_limits}

While symbolic execution has many applications (\eg bug detection, security analysis, debugging, and program repair), it also has well-known limitations, as we briefly summarize below:

\subsubsection{Limitation: Handling Loops (and Recursion)}\label{sec:limit_loop}

Unbounded loops (and recursion) are a known problem for symbolic execution.
Here, symbolic state {\em forking} (\autoref{sec:symexe_intro}) can treat each loop iteration $\{0,1,2,...\}$ as a new path, potentially leading to infinite unrolling if the loop is not bounded.
Popular symbolic execution tools, such as KLEE~\cite{Cadar08}, 
handle this problem using a {\em concrete iteration bound}---replacing potentially infinite exploration with a bounded, but incomplete, exploration.

An alternative is to use {\em loop invariants}.
If known, the invariant allows symbolic execution to pass over a loop without explicit unrolling.
Loop invariants can be manually provided, or discovered automatically, such as using~{\em abstract interpretation}~\cite{Cousot77} over some known domain, {\em constraint based} (CBMC~\cite{gadelha17}), or machine learning (\textsc{Code2Inv}~\cite{Si18}).
However, loop invariant discovery may only work for simple loops, and the general case is either computationally hard or undecidable.

\subsubsection{Limitation: Handling External Environment}\label{sec:limit_env}

Another known problem is the handling of external functions (\eg calls to third-party libraries without source code) and/or external inputs (\eg \verb+recv+ from a socket), collectively the {\em external environment}.
Since the underlying symbolic-execution theorem prover uses {\em deductive reasoning}, a precise specification of all external operations and/or inputs is usually required.
As such, the environment is usually handled through a combination of stubs, modeling, or {\em concretization}.
For example, the user can manually model an external function call by implementing a replacement {\em stub} function that specifies the necessary specification using \verb+klee_assume()+.
However, this approach is manual, and modeling arbitrary code or inputs can require significant effort, meaning that the approach tends to rarely scale.

Another approach is concretization, where the symbolic execution algorithm assigns concrete values to some symbolic variables, allowing external functions to be executed with these values.
However, concretization can also lead to an incomplete exploration of program behavior.

\subsubsection{Limitation: Heap Manipulating Programs}\label{sec:limit_heap}

Traditional theorem provers and SMT solvers tend to have limited support for reasoning over (mutable) data structures with complex structural invariants, such as {\em singly}- and {\em doubly}-linked lists, binary trees, {\em red-black} trees, and DAGs.
By extension, traditional symbolic execution tools inherit these limitations.
Some tools are based on {\em Separation Logic}~\cite{Reynolds02}, which does support reasoning over heap-manipulating programs, as used by VeriFast~\cite{Jacobs11} and Infer~\cite{Distefano19}.
However, VeriFast is manual and annotation-heavy, and Infer uses heuristics to infer common structural invariants, such as linked lists, but not arbitrary heap shapes.

\subsubsection*{Discussion}
\autoref{fig:motivate} (a) is an example of a program that exhibits all three limitations, including unbounded loops (\verb+while (true) ...+), interaction with the external environment (\verb+recv+/\verb+send+), and is a heap manipulating program (\verb+NODE+).
Although this program is relatively simple, it still presents a significant challenge for traditional symbolic execution tools such as KLEE.

\subsection{Limitations of LLMs}\label{sec:llm_limits}

LLMs are trained on a huge corpus of data, and do not necessarily have the same limitations as symbolic execution.
For example, an LLM can reason over 
\autoref{fig:motivate} (a), and can answer queries about the program.
That said, LLMs also have known limitations,
as summarized below:

\subsubsection{Limitation: Scale}\label{sec:limit_scale}

LLMs typically have a limited ability to reason over large and complex code bases.
For example, the {\em token limit} imposes a maximum number of tokens for the given input (and output), and a medium-to-large code base can easily exceed this limit.

\subsubsection{Limitation: Approximate Oracles}\label{sec:limit_oracle}

LLMs rely on {\em approximate reasoning}, \ie answers based on statistical patterns and heuristic generalization, rather than the formal deductive reasoning of traditional theorem provers.
Even if the token limit (\ref{sec:limit_scale}) is not exceeded, studies have shown that LLMs generally perform worse with overly verbose prompts that include irrelevant information~\cite{Levy24}.

\subsubsection*{Discussion}

Although scale and accuracy are concerns, LLMs can typically handle more classes of programs than traditional analysis methods, such as symbolic execution.
Furthermore, the relative completeness of LLMs, in the absence of precise specifications, means that LLMs can be easily applied to a wide range of applications.
The completeness/applicability can often be of greater pragmatic interest than perfect accuracy, and this is one reason behind the explosion of real-world applications.

\subsection{Our Approach}\label{sec:approach}

Many limitations of traditional symbolic execution stem from those of the underlying theorem prover.
Specifically, existing theorem provers and SMT solvers only accept queries in some formal input language, with limited expressiveness compared to the original source language (\eg \verb+C+/Java/Python).
For example, an SMT solver will only accept queries in the form of a Boolean formula over a given set of theories (T), such as linear inequalities, bit vectors, or arrays.
In contrast, the \verb+C+ programming language is significantly more expressive, with complex control-flow (loops), memory, pointers, library calls, data structures, environmental interactions, etc.
The discrepancy in the expressiveness complicates the translation from the high-level programming language into the solver input language---such as ``unrolling'' loops into flat, quantifier-free formulas---and such translation may also be incomplete (\eg unbounded loops).
Furthermore, the SMT solver may not support the necessary theories for reasoning over complex programs, such as heap-manipulating data structures and external environment interactions.

Our underlying approach is to use the path-based decomposition of symbolic execution, but to replace the traditional theorem prover with an LLM.
The key advantage of LLMs is that they can {\em reason over the source code directly}---eliminating the need for translation into a less expressive solver input language and the associated limitations.
Instead, our approach represents the path constraint as a generic {\em strongest post-condition} $\spc(S, \PP)$ predicate transformer~\cite{Dijkstra75} over a pre-condition $\PP$, and a {\em derived sub-program} $S$ which represents a path or set of paths.
Our key insight is that LLMs can reason over $\spc$-constraints directly, since $S$ is just ordinary source code, without the need for ``translation''.
Essentially, we can view the LLM as an effective solver for untranslated $\spc$-constraints ``as-is''.
We demonstrate this concept with a simple example.

\begin{example}[Simple]
Consider the simple program ($\itec{\texttt{x} > \texttt{y}}{\texttt{z} := \texttt{x} + 2}{\texttt{z} := \texttt{x} * \texttt{y}}$) and post-condition $\QQ = (\texttt{z} > \texttt{y})$ once more (see~\autoref{ex:simple}).
Then the path constraint for the $\mathit{else}$-branch can be represented as a formula ($\varphi$) or a sub-program ($S$), as follows:
\vspace{0.3em}
\begin{center}
\begin{tabular}{c|c}
{\em Formula} ($\varphi$) & {\em Strongest Post-Condition over a Sub-program} ($S$) \\
\hline
\cellcolor{cyan!10}
$\quad \neg(\texttt{x} > \texttt{y}) \land \texttt{z} = \texttt{x} \times \texttt{y} \quad$ &
\cellcolor{magenta!10}
$\spc(S, \mathit{true}) \qquad \text{where}~S = \{\assertc{\texttt{x} \leq \texttt{y}}; \texttt{z} := \texttt{x} * \texttt{y}\}$ \\
\end{tabular}
\end{center}
\vspace{0.3em}
The key idea is that both ($\varphi$) and $\spc(S,\mathit{true})$ are {\bf equivalent} representations of the same path constraint---\ie one can be derived from the other under the definition of the language semantics ($\spc$).
While the formula is suitable for a theorem prover, the sub-program is suitable for an LLM:
\begin{align}
\text{``{\em Given $\{\assertc{\texttt{x} \leq \texttt{y}}; \texttt{z} := \texttt{x} * \texttt{y}\}$, does the post-condition $\texttt{z} > \texttt{y}$ always hold?}''}
    \label{eq:vc2}
    \tag{\sc VC-Else-2}
\end{align}
The LLM determines the VC does {\bf not} hold, \eg the following GPT4~\cite{Gpt4} output (emphasis original):
\begin{quote}
``{\em The post-condition $\texttt{z} > \texttt{y}$ {\bf does not always hold}. A simple counterexample is when $\texttt{x}=1$, where $\texttt{z}=\texttt{y}$ instead of $\texttt{z}>\texttt{y}$. Hence, the claim is {\bf false}.}'' $\qquad \qed$
\end{quote}
\end{example}\noindent

\noindent
Our approach is to enumerate all sub-programs based on a {\em partitioning} of paths through the original program. 
Here, each sub-program is algorithmically derived from the original code, and contains all statements that (1) are visited by any path from the partition, and (2) of which post-condition $\QQ$ is data- or control-flow dependent.
Like traditional symbolic execution, our approach is a path-based decomposition of the original program analysis problem into smaller (more tractable) sub-problems. 
This decomposition helps to address some of the limitations of direct LLM-based reasoning (\autoref{sec:llm_limits}), such as scale and accuracy.
Furthermore, since our approach uses an LLM, it avoids many of the limitations of traditional symbolic execution (\autoref{sec:se_limits}) caused by translation. 
We summarize the benefits as follows:

\begin{itemize}[leftmargin=*]
\item[$\rhd$]\hyperref[sec:limit_loop]{\em Handling Loops (and Recursion).}
Our approach avoids translation of (unbounded) into a less expressive language, and loops/recursion can be represented ``as-is'' in the derived sub-program(s).
Similarly, our approach does not need explicit loop invariant recovery or annotation, as  LLMs are capable of reasoning over loops without any special intervention.

\item[$\rhd$]\hyperref[sec:limit_env]{\em Handling External Environment.}
Rather than manual modeling or concretization, our approach is to use the LLM to infer the likely behavior of the environment or external function call.
Since LLMs are trained on a huge corpus of real-world code, they have significant exposure to common libraries, file formats, protocols, etc.
Furthermore, even if the external environment is novel, LLMs can still infer the most likely behavior based on clues from the context (function names, variable names, code comments, placement within the algorithm, etc.), as a form of {\em abductive reasoning}, or {\em inference to the best explanation}, without the need for explicit modeling.

\item[$\rhd$]\hyperref[sec:limit_heap]{\em Heap Manipulating Programs.}
LLMs can directly interpret heap-manipulating programs without the need 
for any special logical framework.
LLMs can also (abductively) infer the (likely) structural invariants based on direct interpretation of the code, without explicit annotation.

\item[$\rhd$]\hyperref[sec:limit_scale]{\em Scale.}
Like traditional symbolic execution, our approach decomposes program analysis problems into smaller (tractable) sub-problems, helping to avoid any hard or soft limit of the LLM.
This allows our approach to scale to large/complex programs and analysis problems.

\item[$\rhd$]\hyperref[sec:limit_oracle]{\em Approximate Oracles.}
Studies~\cite{Levy24} show that LLMs perform better with more targeted and concise prompts.
By decomposing program analysis problems into sub-problems that capture only the relevant parts of the original (possibly large) code base, we help to focus the LLM and improve the overall accuracy of the analysis.
\end{itemize}
The decomposition and lack of translation mean that our approach can handle programs that are difficult for traditional program analysis.
We illustrate with an example.

\begin{example}[LLM-based Symbolic Execution]\label{ex:motive}
Consider the \autoref{fig:motivate} (a) program that cannot easily be handled by traditional symbolic execution methods (\autoref{sec:se_limits}),
and a simple program analysis problem that verifies each \verb+lock(&mutx)+ operation is paired with an \verb+unlock(&mutx)+ operation.
Furthermore, assume that (for the sake of example) the (a) program is too complex for an LLM to handle directly (\autoref{sec:llm_limits}).\footnote{This is not necessarily true, but is an assumption for the sake of an example that can fit within the page limit.}
We can express this as a natural language pre- and post-condition ($\PP$ and $\QQ$ respectively) that ``\verb+mutx+ is unlocked''.
Then \autoref{fig:motivate} (b) is an example of a {\em derived sub-program} that acts as a substitute for the original program with respect to $\PP$ and $\QQ$.
We have that: 
\begin{align*}
    \spc(\text{{\bf (b)}},\PP) \not\models \QQ
    \qquad \Rightarrow \qquad
    \spc(\text{{\bf (a)}},\PP) \not\models \QQ
\end{align*}
Thus, to refute $\QQ$ for (a), it is sufficient to refute $\QQ$ for (b).
Furthermore, program (b) is targeted to the specific program analysis task (that \verb+mutx+ is unlocked), and is ${\sim}80\%$ smaller in token count (419 vs 81).
The LLM determines the post-condition {\bf holds} for (b).

Another example is \autoref{fig:motivate} (c) and the post-condition ($\texttt{db->key} \neq \texttt{NULL}$).
Assuming the same condition initially holds, then (c) is ${\sim}64\%$ smaller (419 vs 149 tokens).
The LLM determines the post-condition does {\bf not hold} for (c) since \verb+strdup()+ may return \verb+NULL+.
$\qed$\end{example}

\subsubsection*{Algorithm}

\newlength{\oldcolumnsep}
\setlength{\oldcolumnsep}{\columnsep}
\setlength{\columnsep}{2em}
\begin{wrapfigure}{r}{0.5\textwidth}{}
\hspace{-1em}
\mbox{\footnotesize
\begin{algorithm}[H]
\SetAlgoLined
\SetAlgoNoEnd
\DontPrintSemicolon
\KwIn{A Hoare triple $\triple{\PP}{C}{\QQ}$}
\KwOut{\texttt{HOLDS} or a counterexample $S$}

\SetKwFunction{SymExe}{LLMSymExe}
\SetKwFunction{Parse}{Parse}
\SetKwFunction{Partition}{Partition}
\SetKwFunction{GenCFG}{GenCFG}
\SetKwFunction{GenPaths}{GenPartitions}
\SetKwFunction{GenSlice}{GenSubProg}
\SetKwFunction{Render}{Render}
\SetKwFunction{LLM}{LLM}
\SetKwProg{Fn}{Function}{:}{}

\BlankLine
\Fn{\SymExe{$\triple{\PP}{C}{\QQ}$}}{
    $\mathit{AST} \leftarrow \Parse{$C$}$\;
    $\mathit{CFG} \leftarrow \GenCFG{$\mathit{AST}$}$\;
    $\mathit{partitions} \leftarrow \GenPaths{$\mathit{CFG}$}$\;
    \For{$\Pi \in \mathit{partitions}$}{
    $S \leftarrow \GenSlice{$\PP, \Pi, \QQ$}$\;
    $\mathit{prompt} \leftarrow \text{\tt "assuming $\PP$"} ~\ensuremath{+\!\!+}$ \;
    \hspace{4.3em} $\Render{$S, \mathit{AST}$} ~\ensuremath{+\!\!+}$ \;
    \hspace{4.3em} $\text{\tt "does $\QQ$ hold?"}$\;
    $\mathit{result} \leftarrow \LLM{$\mathit{prompt}$}$\;
    \lIf{$\mathit{result} = \texttt{FALSE}$}{
        \Return $S$
    }
    }
    \Return \texttt{HOLDS}
}
\BlankLine
\BlankLine

\caption{\small LLM-based Symbolic Execution \label{alg:symexe}}
\end{algorithm}
}
\vspace{-1em}
\end{wrapfigure}

An overview of our LLM-based symbolic execution algorithm is summarized in \autoref{alg:symexe}.
Here, the algorithm's {\em frontend} is similar to that of a standard compiler, and parses the program into an {\em Abstract Syntax Tree} (AST, line 2), and then generates a {\em Control Flow Graph} (CFG, line 3).
Next, the algorithm generates a representation of the set of all {\em paths} through the CFG (line 4).
Here, the set of all paths is represented as a set of path $\mathit{partitions}$ such that ($\mathit{paths} = \Pi_1 \cup ... \cup \Pi_n$), where each partition $\Pi_i$ represents some (possibly infinite) subset of paths.
One challenge is how to generate a {\em good} set of partitioning (to be discussed in \autoref{sec:principles}).
Next, for each partition $\Pi$, the algorithm generates a derived sub-program $S$ that generalizes the partition (line 6). 
For example, \autoref{fig:motivate} (b) and (c) are possible derived sub-programs of \autoref{fig:motivate} (a).
Each sub-program $S$ is used to construct a corresponding prompt (line 7), including {\em rendering} the $S$ back into a text-based source-code representation (line 8).
The prompt (line 7) is a natural language representation of the Hoare triple $\triple{\PP}{S}{\QQ}$, which holds iff $\spc(S,\PP) \models \QQ$.
Finally, the prompt is sent to the LLM  for inference (line 10).
\autoref{alg:symexe} can either establish or refute the post-condition, subject to the reasoning capabilities of the underlying LLM.
Assuming that the partitions are ordered based on slice, \autoref{alg:symexe} will return the {\em least} sub-program $S$ that is deemed to refute the post-condition.
Otherwise, if no such refutation is found, \autoref{alg:symexe} deems that the triple holds (\verb+HOLDS+).
\
\setlength{\columnsep}{\oldcolumnsep}
\begin{table}
{\scriptsize
\caption{Summary of main similarities and differences between traditional and LLM-based symbolic execution.
Here ({\em Cap} = {\em Capabilities}), ({\em Imp} = {\em Implementation}), ($\spc$ = {\em strongest post-condition}), and (\colorbox{cyan!20}{the key differences}).\label{fig:compare}}
\begin{tabular}{l|l|p{0.36\textwidth}|p{0.35\textwidth}}
\toprule
& {\em Property} & {\em Tradition Symbolic Execution} & {\em LLM-based Symbolic Execution} \\
\hline
\hline
\multirow{6}{*}{\rotatebox{90}{\em Design}} & {\em Overall method} & Decomposition \& solving path constraints & Decomposition \& solving path constraints \\
& \cellcolor{cyan!20} {\em Decomposition method} & \cellcolor{cyan!20} Decomposition into a formal language & \cellcolor{cyan!20} Decomposition into sub-programs \\
& \cellcolor{cyan!20} {\em Reasoning engine} & \cellcolor{cyan!20} Theorem prover or SMT solver & \cellcolor{cyan!20} Large Language Model (LLM) \\
& \cellcolor{cyan!20} {\em Reasoning method} & \cellcolor{cyan!20} Deductive & \cellcolor{cyan!20} Approximate, deductive, abductive \\
& \cellcolor{cyan!20} {\em Path representation} & \cellcolor{cyan!20} Formal language (unfolded $\spc$-path constraints) & \cellcolor{cyan!20} Untranslated $\spc$-constraints over truncated slices \\ 
& {\em Specification language} & Formal language & Either formal, code, or natural language \\
\hline
\multirow{4}{*}{\rotatebox{90}{\em Cap.}} & 
{\em Weaknesses} & Loops, environment, heaps & Complex integer, linear, Boolean reasoning \\
& {\em Unbounded loops} & Infinite unfolding or loop invariants & LLM reasons over loops ``as-is'' \\
& {\em Environment} & Manual modeling/specifications required & LLM abductive reasoning or training data \\
& {\em Heap manipulation} & Manual annotation + Separation Logic & LLM reasons over heaps ``as-is'' \\
\hline
\multirow{2}{*}{\rotatebox{90}{\em Imp.}} & {\em Programming languages} & Language specific (\texttt{C}+KLEE~\cite{Cadar08} and Java+SPF~\cite{Spf10}) & Programming language agnostic \\
& {\em Compiler infrastructure} & Close integration (LLVM~\cite{Chris04}+KLEE, etc.) & Lightweight (AST-level) implementation \\
\bottomrule
\end{tabular}
}
\setlength{\fboxsep}{0pt}
\end{table}

\subsubsection*{Summary}
Like traditional symbolic execution, LLM-based symbolic execution (\autoref{alg:symexe}) represents a path-based {\em decomposition} the original program analysis task into smaller subtasks. 
A summary of the main similarities and differences is shown in \autoref{fig:compare}.
The substitution of a theorem prover with an LLM changes to various aspects of the design, capabilities, and implementation of the symbolic execution engine.
For example, LLMs use {\em approximate} and {\em abductive} reasoning, or rely on information learned during the training process, meaning that LLMs do not need precise specifications or environment modeling.
Likewise, the lack of translation into some (less expressive) formal language allows the LLM to reason over loops or heap manipulation ``as-is'', without relying on loop/data-structure invariant discovery.

In this paper, we study the concept of program analysis via LLM-based symbolic execution.
First, we study the {\em principles} of LLM-based symbolic execution in terms of 
the idealized procedural programming language used by Hoare logic~\cite{Hoare69}.
We show that program analysis tasks can be decomposed into tasks over derived sub-programs representing paths, or sets of paths (truncated slices), through the original program.
Furthermore, we also show that only a {\em finite} number of $\mathit{partitions}$ (\autoref{alg:symexe}, line 4) needs to be considered, ensuring that LLM-based symbolic execution will always terminate---even for unbounded loops.

In addition, we study the application of LLM-based symbolic execution in {\em practice}.
For this, we design \tool---an LLM-based symbolic execution engine for real-world programming languages such as \verb+C+/Java/Python.
Our approach is based on the observation that \autoref{alg:symexe} is mostly language agnostic except for specific aspects, such as the parser.
This means our approach can be readily ported to other programming languages.
Furthermore, we also observe that, since LLMs are fundamentally approximate, we can build a {\em lightweight} implementation that uses {\em approximate} parsing and dependency analysis---without relying on any heavyweight and/or language-specific compiler framework.

\section{Principles of LLM-Based Symbolic Execution}\label{sec:principles}

Our goal is to adapt traditional symbolic execution methods to LLMs that reason over code directly, rather than translation into a (less expressive) theorem prover input language.

\subsection{Symbolic Execution Foundations}\label{sec:symexe}

We use the {\em minimal imperative language} defined by Hoare logic~\cite{Hoare69} augmented with an explicit $\mathbf{assume}$-statement.\footnote{Note $\assertc{B}$ can be emulated as ($\whilec{\neg B}{\skipc}$) under partial correctness, but is treated as a special case.}
We define the language syntax as follows:
{
\begin{align*}
C &::= \skipc 
  \mid C; C 
  \mid \assertc{B}  
  \mid x := E 
  \mid \itec{B}{C}{C}
  \mid \whilec{B}{C}
\end{align*}
}\noindent
Where $E$ represents some base language (\eg arithmetic expressions) and $B$ represents Boolean expressions over $E$.
We also use $\epsilon$ to sometimes represent an empty program (equivalent to $\skipc$).
The language semantics are defined inductively (\ie least relation) in terms of the {\em strongest post-condition} ($\spc$) relation defined as follows:

\vspace{-1em}
{
\begin{align*}
\qquad \spc(\skipc,\PP) = \PP \quad \spc(\{C_1; C_2\}, \PP) & = \spc(C_2, \spc(C_1, \PP)) \quad \spc(\assertc{b},\PP) = b \land \PP\\
\spc(x := e, \PP) & = \exists y : x = e[x \mapsto y] \land \PP[x \mapsto y] \\
\spc(\itec{b}{C_1}{C_2}, \PP) & = \spc(C_1, b \land \PP) \lor \spc(C_2, \neg b \land \PP) \\
\spc(\whilec{b}{C}, \PP) & = \spc(C;\whilec{b}{C}, b \land \PP) \lor (\neg b \land \PP) 
\end{align*}
}

\noindent
We define a {\em linear program} to be any program comprising only $\skipc$-, $\mathbf{assume}$-, and {\em assignment}-statements (without conditionals or loops).
We define a $\mathit{path}$ to be a linear program that is derived by unfolding the original program $C$ using the following rules:

\vspace{-1em}
{
\small
\begin{align*}
\unfold(\skipc,\pp) = \{\pp; \skipc\} \qquad
\unfold(\assertc{b},\pp) & = \{\pp; \assertc{b}\} \qquad
\unfold(x := e,\pp) = \{\pp; x := e\} \\
\unfold(\{C_1; C_2\},\pp) & = \cup \{\unfold(C_2,\pp') ~|~ \pp' \in \unfold(C_1,\pp)\} \\
\unfold(\itec{b}{C_1}{C_2},\pp) & = \cup
\begin{cases}
\unfold(C_1,\{\pp;\assertc{b}\}) \\
\unfold(C_2,\{\pp;\assertc{\neg b}\}) \\
\end{cases} \\
\unfold(\whilec{b}{C},\pp) & =
\cup
\begin{cases}
\unfold(\{C; \whilec{b}{C}\},\{\pp;\assertc{b}\}) \\
\{\pp;\assertc{\neg b}\} \\
\end{cases}
\end{align*}
}\noindent
We abstractly define {\em symbolic execution} to be any algorithm that combines {\em path unfolding} with {\em Verification Condition} (VC) {\em solving}.
Given a program analysis task represented as a Hoare~\cite{Hoare69} triple $\triple{\PP}{C}{\QQ}$, then a {\em symbolic execution} algorithm: 
\begin{enumerate}[leftmargin=*]
    \item exhaustively {\em generates} the set of all $\mathit{paths} = \unfold(C,\epsilon)$ through the program; and
    \item \label{case:solve} {\em solves} each corresponding VC ($\spc(\mathit{path},\PP) \models \QQ$) for each $\mathit{path} \in \mathit{paths}$.
\end{enumerate}
That is, a symbolic execution algorithm computes the following using explicit enumeration:

\vspace{-1em}
{\small
\begin{align}\label{eq:symexe}
\triple{\PP}{C}{\QQ} \qquad \text{iff} \qquad
\bigwedge \{\spc(\pi, \PP) \models \QQ ~|~\pi \in \unfold(C,\epsilon) \}
\tag{\sc SymExe}
\end{align}
}

\noindent
The triple is deemed to {\em hold} if each individual VC holds for the corresponding path ($\pp$), and to {\em not hold} otherwise.
Here, each $\spc(\pi,\PP)$ is defined to be the {\em path condition} for the corresponding path $\pp$.
Essentially, a symbolic execution algorithm decomposes the original program analysis task into simpler subtasks that can be solved separately.

\subsubsection*{Traditional Symbolic Execution}

``Traditional'' symbolic execution algorithms solve each VC using a suitable {\em theorem prover}, such as an SMT solver~\cite{DeMoura08}.
To do so, the $\spc$-constraints for each path are {\em translated} into a logical formula $\varphi$ over the domain of $E$, by applying the (linear program subset of the) $\spc$-rules defined above.
The translated VC ($\varphi \models \QQ$) is then solved using a theorem prover.
Note that the translation is necessary since traditional theorem provers are limited to a specific input language (\eg SMT-LIB), and cannot make inferences over an abstract $\spc$-constraint directly.
In addition, most practical symbolic execution tools implement several optimizations, including incremental path unfolding, incremental $\spc$-translation, {\em pruning} infeasible paths, and the {\em merging} of similar paths.
For example, rather than pre-computing the set of paths upfront, most practical implementations maintain a \emph{symbolic state} comprising a current location, a (partially constructed) path constraint/condition, and a set of current variables-of-interest.
Such optimizations are consistent with abstract symbolic execution (\autoref{eq:symexe}) defined above.

We can also understand the limitations of \autoref{sec:se_limits}.
Firstly, the number of paths in the set $\unfold(C,\epsilon)$ may be very large (exponential) or even infinite (unbounded loops, \autoref{sec:limit_loop}).
This is known as the {\em path explosion} problem, and is a well-known limitation of symbolic execution methods.
Secondly, the $\spc$-translation will be limited for programs that contain operations that are not supported---such as common interactions with the external environment (\autoref{sec:limit_env}).
Finally, the underlying theorem prover itself may be limited.
For example, KLEE~\cite{Cadar08} uses the Z3~\cite{Baldoni18} SMT solver over the domain of linear arithmetic, arrays, and bit-vectors by default.
However, this configuration does not support reasoning over heap-manipulating programs (\autoref{sec:limit_heap}).

\begin{figure}
\centering
\renewcommand{\arraystretch}{0.5}
\begin{tabular}{ccc}
{\footnotesize \textcolor{red!60!black}{(a)} simple program ($C_\mathit{simple}$)} &
{\footnotesize \textcolor{red!60!black}{(b)} path ($\pp_\mathit{simple}$)} &
{\footnotesize \textcolor{red!60!black}{(c)} truncation ($T_\mathit{simple}$)} \\
\fcolorbox{black}{yellow!14}{
\begin{minipage}[t]{0.26\textwidth}
{
\footnotesize
\setstretch{0.7}
\begin{algorithm}[H]
\texttt{i} $:= 1$\;
\While{\texttt{i} $\leq$ \texttt{n}} {
    $\mathbf{read}(\texttt{x})$\;
    \uIf{\texttt{x} $< 0$ }{
        \texttt{xs}.$\mathit{delete}(-\texttt{x})$\;
    }
    \uElse{
        \texttt{xs}.$\mathit{insert}(\texttt{x})$\;
    }
    \texttt{z} $:= \texttt{xs}.\mathit{size}()$\;
    $\mathbf{write}(\texttt{z})$\;
    \texttt{i} $:= \texttt{i} + 1$\;
}
\end{algorithm}
}
\end{minipage}
}
&
\fcolorbox{black}{red!7}{
\begin{minipage}[t]{0.26\textwidth}
{\footnotesize
\setstretch{0.7}
\begin{algorithm}[H]
\texttt{i} $:= 1$\;
$\assertc{\texttt{i} \leq \texttt{n}}$\;
$\mathbf{read}(\texttt{x})$\;
$\assertc{\neg(\texttt{x} < 0)}$\;
\texttt{xs}.$\mathit{insert}(\texttt{x})$\;
\texttt{z} $:= \texttt{xs}.\mathit{size}()$\;
$\mathbf{write}(\texttt{z})$\;
\texttt{i} $:= \texttt{i} + 1$\;
$\assertc{\texttt{i} \leq \texttt{n}}$\;
$\mathbf{read}(\texttt{x})$\;
$\assertc{\neg(\texttt{x} < 0)}$\;
\texttt{xs}.$\mathit{insert}(\texttt{x})$\;
\texttt{z} $:= \texttt{xs}.\mathit{size}()$\;
$\mathbf{write}(\texttt{z})$\;
\texttt{i} $:= \texttt{i} + 1$\;
$\assertc{\neg(\texttt{i} \leq \texttt{n})}$\;
\end{algorithm}
}
\end{minipage}
}
&
\parbox[t]{0.26\textwidth}{
\centering
\fcolorbox{black}{green!7}{
\begin{minipage}[t]{0.26\textwidth}
{
\footnotesize
\setstretch{0.7}
\begin{algorithm}[H]
\texttt{i} $:= 1$\;
\While{\texttt{i} $\leq$ \texttt{n}} {
    $\mathbf{read}(\texttt{x})$\;
    $\assertc{\neg(\texttt{x} < 0)}$\;
    \texttt{xs}.$\mathit{insert}(\texttt{x})$\;
    \texttt{z} $:= \texttt{xs}.\mathit{size}()$\;
    $\mathbf{write}(\texttt{z})$\;
    \texttt{i} $:= \texttt{i} + 1$\;
}
\end{algorithm}
}
\end{minipage}
} \\
\vspace{0.5em}
{\footnotesize \textcolor{red!60!black}{(d)} slice ($S_\mathit{base})$} \\
\vspace{-0.5em}
\fcolorbox{black}{green!7}{
\begin{minipage}[t]{0.26\textwidth}
{
\footnotesize
\setstretch{0.7}
\begin{algorithm}[H]
\texttt{i} $:= 1$\;
\While{\texttt{i} $\leq$ \texttt{n}} {
    $\mathbf{read}(\texttt{x})$\;
    \texttt{xs}.$\mathit{insert}(\texttt{x})$\;
    \texttt{i} $:= \texttt{i} + 1$\;
}
\end{algorithm}
}
\end{minipage}
}
}
\end{tabular}
\caption{\textcolor{red!60!black}{(a)} A simple example program ($C_\mathit{simple}$), \textcolor{red!60!black}{(b)} one example path ($\pp_\mathit{simple}$) through ($C_\mathit{simple}$), \textcolor{red!60!black}{(c)} a truncation ($T_\mathit{simple}$) of ($C_\mathit{simple}$) assuming only the inner $then$-branch is taken, and \textcolor{red!60!black}{(d)}
a slice ($S_\mathit{simple}$) of ($T_\mathit{simple}$) assuming $\{\texttt{n},\texttt{xs}\}$ are the vars-of-interest.
\label{fig:example}}
\end{figure}

\subsection{LLM-Based Symbolic Execution (Path-Based)}\label{sec:path}
The basic idea behind ``LLM-based'' symbolic execution is to use a {\em Large Language Model} (LLM) as the underlying reasoning engine instead of a theorem prover.
Thus, given a VC of the form ($\spc(\pp, \PP) \models \QQ$), we can use the LLM to reason directly over the path constraint and post-condition $\QQ$.
This is possible since, through its training process, the LLM can interpret both the syntax of the path $\pp$ (represented as ordinary code), as well as the language semantics represented by the $\spc$-rules.
We illustrate with the following example.

\begin{example}[Path-based LLM-based Symbolic Execution]\label{ex:path}
We consider the simple example shown in \autoref{fig:example}~(a), which is based on Example Program 3 in~\cite{Bogdan88}~.
We assume that $\texttt{xs}$ is a data structure implementing a multi-set, and $\mathit{insert}$/$\mathit{delete}$/$\mathit{size}$ can be expanded to sub-programs with a suitable multi-set implementation (\eg a singly linked list), and is initially empty.
Furthermore, we assume that the ($\mathbf{read}$) operation always returns a positive number, which can be expressed in natural language or as a formal rule ($\forall \texttt{x}, \mathcal{R}: \spc(\mathbf{read}(\texttt{x}),\mathcal{R}) \rightarrow \texttt{x} > 0$).
Under these assumptions, the final size of $\texttt{xs}$ should be equal to $\texttt{n}$.
We can express this as the following triple:

\vspace{-1em}
{
\small
\begin{align}
\triple{\texttt{n} \geq 0 \land \texttt{xs}.\mathit{size}() = 0 \land \text{``$\textbf{read}(\texttt{x})$ always returns a positive number''}}{~C_\mathit{simple}~}{\texttt{xs}.\mathit{size}()=\texttt{n}}
\tag{\textsc{Triple}}\label{eq:example}
\end{align}
}

\noindent
Although conceptually simple, the \autoref{fig:example}~(a) program is challenging for several reasons, including an unbounded loop, environmental input ($\mathbf{read}$), and data-structure reasoning ($\texttt{xs}$).
We may prove (\autoref{eq:example}) by enumerating paths, such as ($\pp_\mathit{simple}$) from \autoref{fig:example}~(b), representing two iterations of the loop.
We can encode the VC ($\spc(\pp_\mathit{simple},\PP) \models \QQ$) as a prompt, which is confirmed using an LLM:
\begin{quote}
    ``{\em Given the pre-condition $\PP$ and the code $\pp_\mathit{simple}$, does the post-condition $\QQ$ hold?}'' $\qed$
\end{quote}
\end{example}

\noindent
This example shows how LLMs can solve tasks that are challenging for traditional program analysis methods.
That said, a simple path-based decomposition still inherits some limitations.
Firstly, 
there can still be an infinite number of paths (\ie path explosion), leading to non-termination.
Secondly, 
each individual path could still be too long for the LLM to effectively reason over (\eg path $\pp_\mathit{simple}$ from \autoref{fig:example} (b) can be generalized to any length).
Finally, paths can accumulate irrelevant statements, such as variable $\texttt{z}$, that can also exacerbate the path length problem.

\subsection{LLM-Based Symbolic Execution (Slice-Based)}\label{sec:slice}

To address the path-explosion problem, our core idea is to merge individual $\spc$-constraints (representing individual paths) into {\em generalized $\spc$-constraints} representing (possibly infinite) {\em sets of paths}.
Given a {\em partition} $\Pi \subseteq \unfold(C,\epsilon)$, our approach constructs a {\em truncated sub-program} $T_\Pi$ such that: 

\vspace{-1em}
{
\begin{align}\label{eq:general}
\spc(\pp,\PP) \models \spc(T_\Pi,\PP) \qquad \text{for all $\pp \in \Pi$} 
\tag{\sc Generalization}
\end{align}
}

\noindent
Thus, instead of disposing of a (possibly infinite) number of verification conditions (VCs) of the form ($\spc(\pp, \PP) \models \QQ$) for each $\pp \in \Pi$, our approach disposes of a single VC of the form ($\spc(T_\Pi,\PP) \models \QQ$) for the entire set ($\Pi$).
First, we shall present a method for constructing a truncated sub-program for a given partition.
Next, we shall present a partitioning algorithm that needs only consider a finite number of subsets, even for programs with infinite paths (unbounded loops), ensuring the symbolic execution algorithm always terminates.

\subsubsection{Construction}\label{sec:construction}
We consider the construction of truncated sub-programs.

\myparagraph{Truncation}
Given a program $C$ and a (possibly infinite) subset of $\Pi \subseteq \unfold(C,\epsilon)$, we derive a {\em truncated sub-program} $T_\Pi$ equivalent to $C$ for all $\pp \in \mathit{paths}$, and is \emph{unreachable} otherwise:

\vspace{-1em}
{
\begin{align}
    \Pi \quad \subseteq \quad \unfold_\mathit{reachable}(T_\Pi,\epsilon) \quad \subseteq \quad \unfold(C,\epsilon)
    \tag{\sc Truncation}\label{eq:derived}
\end{align}
}

\noindent
Here, ($\unfold_\mathit{reachable}$) excludes all paths that terminate abnormally via a special $\assumec{0}$ (\ie~{\em assume false}) statement.
Thus, $\assumec{0}$ represents a statement that is assumed to be {\em unreachable}.\footnote{Similar to \texttt{\_\_builtin\_unreachable()} from \texttt{gcc}.}
Consider all statements ($s$) in $C$ that are {\em not covered} by any path $\pp \in \Pi$, then we can construct $T_\Pi$ by replacing all such ($s$) with $\assumec{0}$.
We can represent this idea using the following rewrite rule:\footnote{We use the standard notation ($\mathit{lhs} \rightarrow \mathit{rhs}$) to mean that any term matching the $\mathit{lhs}$ is rewritten to the term matching $\mathit{rhs}$.}

\vspace{-1em}
{
\begin{align}
s \rightarrow \assumec{0} \qquad \text{if $s \not\in \pp$ for all $\pp \in \mathit{paths}$}
\tag{\sc Unreachable}
\end{align}
}

\noindent
We may also apply the following rules to further simplify the resulting sub-program:

\vspace{-1em}
{\small
\begin{align*}
\assumec{0}; C_2 \rightarrow \assumec{0} \qquad C_1; \assumec{0} \rightarrow \assumec{0} \qquad \qquad \qquad \qquad \\
\itec{b}{C_1}{\assumec{0}} \rightarrow \assumec{b}; C_1 \qquad
\itec{b}{\assumec{0}}{C_2} \rightarrow \assumec{\neg b}; C_2 \\\whilec{b}{\assumec{0}} \rightarrow \assumec{\neg b} \qquad \qquad \qquad \qquad \qquad \qquad
\end{align*}
}

\noindent
These rules preserve the (\autoref{eq:derived}) property while also reducing the size (token count) of the resulting $T_\Pi$, which is beneficial for LLM prompting.

\myparagraph{Slicing}
We can further reduce the size of the truncated sub-program using {\em program slicing}~\cite{Weiser81}.
We define a {\em slice} ($S_\Pi$) to be a sub-program derived from $T_\Pi$ by {\em deleting} (\ie replacing by $\epsilon$) any statement ($s$) in $T_\Pi$ that does not violate the condition:

\vspace{-1em}
{
\begin{align}
    \spc(T_\Pi,\PP) \models \QQ \qquad \text{iff} \qquad \spc(S_\Pi,\PP) \models \QQ
    \tag{\sc Slice}\label{eq:slice}
\end{align}
}

\noindent
Our main result is as follows.
If ($\spc(S_\Pi,\PP) \models \QQ$) holds, then:
\begin{enumerate}
\item \label{case:trunc} ($\spc(T_\Pi,\PP) \models \QQ$) holds by (\autoref{eq:slice}); and
\item ($\spc(\pp,\PP) \models \QQ$) for all $\pp \in \Pi$ holds by (\ref{case:trunc}) and (\autoref{eq:general}).
\end{enumerate}
In other words, the single VC $(\spc(S_\Pi,\PP) \models \QQ)$ is sufficient to dispose of the entire partition ($\Pi$).
We can apply standard slicing methods, such as Weiser's back slicing algorithm~\cite{Weiser81} (with $\vars{\QQ}$ as the slice criterion), to construct $S_\Pi$ from $T_\Pi$.
The back slicing algorithm traverses the {\em Control Flow Graph} (CFG), and deletes any statement that is not data- or control-flow-dependent on $\vars{\QQ}$.
The slicing algorithm is illustrated in \autoref{alg:slice}.

\begin{example}[Slice-based Symbolic Execution]\label{ex:slice}
We consider \autoref{ex:path} once more.
Here, the truncated slice ($S_\mathit{simple}$) in \autoref{fig:example}~(d) is a generalization of the infinite partition ($\Pi$) representing {\em all} feasible paths through the $\mathbf{while}$-loop, including
the \autoref{fig:example} (b) path ($\pp_\mathit{simple}$).
An LLM can be used to verify the generalized {\em verification condition} (VC) encoded in natural language.
\begin{quote}
    ``{\em Given the pre-condition $\PP$ and the code $S_\mathit{simple}$, does the post-condition $\QQ$ hold?}''
\end{quote}
Path-based symbolic execution will infinitely unroll the $\mathbf{while}$-loop, generating a new VC for each loop iteration.
In contrast, slice-based symbolic execution requires only a single VC to be checked.
Furthermore, ($S_\mathit{simple}$, 5 lines, 28 tokens) is simpler and more compact than all of ($C_\mathit{simple}$, 10 lines, 56 tokens), ($\pp_\mathit{simple}$, 16 lines, 85 tokens), and ($T_\mathit{simple}$, 8 lines, 48 tokens).
Truncation and slicing are useful for removing irrelevant statements while preserving paths, meaning that the corresponding prompt is simpler and more concise, thereby improving the accuracy of LLMs.
$\qed$\end{example}

\setlength{\oldcolumnsep}{\columnsep}
\setlength{\columnsep}{2em}
\begin{wrapfigure}{r}{0.5\textwidth}{}
\hspace{-1em}
\mbox{\footnotesize
\begin{algorithm}[H]
\SetAlgoLined
\DontPrintSemicolon
\SetKwInOut{KwGlobal}{Globals}
\KwIn{A CFG sub-graph $G$; reverse topological order}
\KwOut{A $\mathit{slice} \subseteq G$}

\SetKwFunction{genSlice}{GenSlice}
\SetKwFunction{getVars}{getVars}
\SetKwFunction{getDependencies}{getDependencies}
\SetKwProg{Fn}{Function}{:}{}

\BlankLine
\Fn{\genSlice{$G$, $\QQ$}}{
    $\mathit{slice} \leftarrow \emptyset$;
    $\mathit{vars} \leftarrow $ \getVars{$\QQ$} \;
    \For{$s \in G$}{
        \uIf{$s$ modifies any $v \in \mathit{vars}$, \textbf{or} \\
            \hspace{.7em} conditional $s$ reads any $v \in \mathit{vars}$}{
            $\mathit{slice} \gets \mathit{slice} \cup \{s\}$\;
            $\mathit{vars} \gets \mathit{vars} \cup \getDependencies(s)$ \;
        }
    }
    \Return{$\mathit{slice}$}\;
}
\BlankLine
\BlankLine

\caption{\small Basic back-slicing algorithm. \label{alg:slice}}
\end{algorithm}
}
\end{wrapfigure}

\myparagraph{Discussion}
Our approach is related to {\em path merging} and {\em loop invariants} in traditional symbolic execution.
Here, given a set of $n$ {\em path constraints}, represented as translated logical formulae $\varphi_i, i \in 1..n$, the idea is to find a formula $\phi$ that is {\em generalization} ($\varphi_i \models \phi$).
Thus, the $n$ verification conditions ($\varphi_i \models \QQ$) can be combined into a single verification condition ($\phi \models \QQ$), helping to mitigate the path explosion problem.
Similarly, the (possibly infinite) set of path constraints $\varrho_i$ through a loop can be generalized into a {\em loop invariant} $I$, such that ($\varrho_i \models I$), allowing the loop to be handled without infinite unrolling.
However, loop-invariant discovery over formulae is difficult and undecidable in the general case.
In contrast, our approach avoids the problem, since the $\spc$-constraints are never translated into a different representation.

\subsubsection{Partitioning}

\begin{wrapfigure}{r}{0.5\textwidth}{}
\hspace{-1em}
\mbox{\footnotesize
\begin{algorithm}[H]
\SetAlgoLined
\SetAlgoNoEnd
\DontPrintSemicolon
\SetKwInOut{KwGlobal}{Globals}
\KwIn{The {\em Control-Flow Graph} (CFG)}
\KwOut{A coverage-based partitioning}
\KwGlobal{Coverage map ($\mathit{covMap}$), end node ($\mathit{end}$)}

\SetKwFunction{DFS}{GenPartitions}
\SetKwProg{Fn}{Function}{:}{}

\BlankLine
\Fn{\DFS{$\mathit{node}, \mathit{pathCov}, \mathit{path}$}}{
    \lIf{$\mathit{pathCov} \in \mathit{covMap}$}{
        \Return $\emptyset$
    }
    $\mathit{pathCov} \leftarrow \mathit{pathCov} \cup \{\mathit{node}\}$\;
    $\mathit{path} \leftarrow \mathit{path} ~\ensuremath{+\!\!+}~ \mathit{node}$ \;
    $\mathit{covMap} \leftarrow \mathit{covMap} \cup \{\mathit{path}\}$ \;
    \lIf{$\mathit{node} = \mathit{end}$}{
        \Return $\{\mathit{path}\}$
    }
    $\mathit{partitions} \leftarrow \emptyset$\;
    \For{$\mathit{succ} \in \text{successors}(\mathit{node})$}{
        $\mathit{partitions} \leftarrow \mathit{partitions}~\cup$ \;
        \qquad \DFS{$\mathit{succ}, \mathit{pathCov}, \mathit{path}$}
    }
    \Return $\mathit{partitions}$
}
\BlankLine

\BlankLine
\BlankLine

\caption{\small Path partitioning algorithm \label{alg:gen}}
\end{algorithm}
}
\end{wrapfigure}

\autoref{sec:construction} describes the construction of a {\em truncated slice} $S$ given a partition ($\Pi$) that subset of paths ($\Pi \subseteq \unfold(C,\epsilon)$).
In principle, any partitioning ($\Pi_1 \cup ... \cup \Pi_n = \unfold(C,\epsilon)$) can be used.
However, a {\em good} partitioning aims to minimize the slice ($S_i$) {\em size} for each $\Pi_i$, $i \in 1..n$, in order to simplify the prompts ultimately sent to the LLM.
Our approach is to construct a partitioning based on {\em path coverage}.

\myparagraph{Path Coverage}
Under \autoref{sec:symexe}, we define a \emph{path} ($\pp$) to be a {\em linear program} over statements or branches (represented as assertions) from $C$.
Here, we define a {\em Control Flow Graph} (CFG) {\em path} to be a sequence of nodes (\aka locations) $\langle l_1, ..., l_m \rangle$ through the CFG representation of $C$.
Given a CFG path ($\pp$), we define {\em path coverage} to be the set ($\cov{\pp} = \{l_1, ..., l_m\}$), \ie the reinterpretation of ($\pp$) from a sequence to a set.
It is common for distinct paths to have the same coverage; for example, different iterations of the same loop have distinct sequences, but produce equivalent sets (\ie covering the same nodes).
We make two key insights.
First, paths with distinct coverage will have distinct truncations, since each path must differ by at least one location.
This will be used as the basis for partitioning.
Secondly, the number of distinct paths w.r.t. coverage is {\em finite}, meaning that LLM-based symbolic execution (slice-based) necessarily terminates---even for unbounded loops.

\myparagraph{Partitioning Algorithm}
The core idea is to enumerate CFG paths ($\pp_\mathit{cov}$) with distinct coverage ($\mathit{cov}$).
Each partition ($\Pi_\mathit{cov} \subseteq \unfold(C,\epsilon)$) is implicitly defined as all paths with the same path coverage as ($\mathit{cov}$).
The algorithm uses a {\em Depth First Search} (DFS) exploration over the CFG, as illustrated in \autoref{alg:gen}.
Here, the algorithm takes a CFG representation of $C$, and generates a {\em partitioning} as a set of {\em representative paths} ($\pp_\mathit{cov}$) for each distinct coverage class ($\mathit{cov}$).
The algorithm works as a standard DFS path construct algorithm, but also maintains a {\em coverage map} ($\mathit{covMap}$) that tracks previously-seen coverage prefixes.
When a ($\mathit{node} \in \mathit{CFG}$) is visited, the coverage map is consulted, and the path is {\em pruned} if the prefix has been observed before.
The coverage map ensures that \autoref{alg:gen} both (1) terminates, and (2) only returns paths (partitions) that differ by at least one location/node.
Each output path ($\pp_\mathit{cov}$) corresponds to a partition ($\Pi_\mathit{cov}$), which is then used to construct a corresponding truncated sub-programs ($T_\mathit{cov}$) and slice ($S_\mathit{cov}$) that is used for LLM prompting.

Given two distinct coverages ($\mathit{cov}$, $\mathit{cov}'$), the resulting slices (\ie $S_\mathit{cov}$, $S_{\mathit{cov}'}$) can sometimes be equivalent ($S_\mathit{cov} = S_{\mathit{cov}'}$) if all distinguishing nodes are removed by the slicer.
Furthermore, some truncated slices may generalize others, \ie ($\spc(S_\mathit{cov},\PP) \models \spc(S_{\mathit{cov}'}, \PP)$).
For example, under \autoref{alg:gen}, then $S = \langle \texttt{i} := 1; \assertc{\neg(\texttt{i} \leq \texttt{n})}\rangle$ is also a valid slice of ($C_\mathit{simple}$, \autoref{fig:example}) representing zero iterations of the $\mathbf{while}$-loop.
Nevertheless, our LLM-based symbolic execution algorithm orders VCs by size, in order to find a {\em least} refutation of the post-condition ($\QQ$).

\myparagraph{Discussion}
\autoref{alg:gen} is guaranteed to {\em terminate}, since the number of possible distinct coverage sets is finite.
Thus, unlike path-based symbolic execution, slice-based symbolic execution will always terminate, even for programs with unbounded loops.
For example, the truncated slice ($S_\mathit{simple}$) from \autoref{fig:example} (b) generalizes infinitely many paths through the loop (for any number of iterations).
Furthermore, the size of each slice $S$ is always bounded by the size of the original program $C$, whereas unrolled paths can exceed any length.

\setlength{\columnsep}{\oldcolumnsep}

\section{LLM-Based Symbolic Execution in Practice}\label{sec:practice}

\begin{figure}[t]
    \centering
    \includegraphics[scale=3.6]{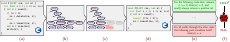}
    \caption{Example workflow.
    Here,
    (a) is the original source code,
    (b) is the {\em Abstract Syntax Tree} (AST) parsed from the code,
    (c) is a generated path slice (\autoref{alg:gen} and \autoref{alg:slice}),
    (d) is the slice {\em rendered} back into the original source code, and
    (e) is the generated {\em prompt} comprising a pre-condition, the slice, and the post-condition, and
    (f) is the LLM inference.
    \label{fig:workflow}}
\end{figure}

The truncated slice-based symbolic execution algorithm of \autoref{sec:principles} is defined for an idealized programming language (the language of Hoare triples).
In this section, our aim is to port our abstract approach to real-world programming languages (\eg \verb+C+/Java/Python).
Our first main insight is that the main symbolic execution and slicing algorithms are {\em agnostic} to the target programming language---provided there is a method for parsing the source code into some suitable representation for path generation and slicing.
Our second main insight is that, since LLMs are {\em approximate oracles}, the greater workflow can also be approximate---meaning that each individual step need not be perfectly precise,
provided the overall accuracy is not significantly degraded.
These insights significantly simplify our overall design.

An example of the workflow is illustrated in \autoref{fig:workflow}.
First, our workflow uses a lightweight parsing framework (specifically, \verb+tree-sitter+~\cite{tree_sitter}) to parse the source code (a)
into an {\em Abstract Syntax Tree} (AST) representation (b).
From the AST, a {\em Control-Flow Graph} (CFG) is constructed, allowing for the finite enumeration of all path partitions using \autoref{alg:gen}.
For each enumerated partition, a {\em truncation} and {\em slice} is constructed (c), by removing statements that are not control- or data-flow dependent on the post-condition $\QQ$, using \autoref{alg:slice}.
Next, the slice is {\em rendered} back into the original source language (d).
Essentially, this means emitting sliced path elements, deleting (\ie not emitting) non-sliced path elements, and replacing (\ie replace by \verb+assume(0)+) all other non-path elements.
The resulting slice generalizes a (possibly infinite) set of paths through the program.
Finally, the slice is used to construct a text-based prompt (e) used to query an LLM (f).
steps (c)-(f) are repeated for each partition using \autoref{alg:symexe}, until exhaustion or the discovery of a counterexample.

\subsection{Partitioning and Truncated Slice Generation}

\begin{figure}
\centering
\includegraphics[scale=4.1]{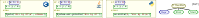}
\caption{Example AST representation of an {\em if-then-else} statement across the \texttt{C}, Java, and Python programming languages.
Each AST node is represented {\em generically} by the {\em file range} from which the element was parsed, as illustrated by the nested colored boxes.
\label{fig:uast}}
\end{figure}

\myparagraph{Parsing}
We use the \verb+tree-sitter+ framework~\cite{tree_sitter} to parse the source code into an AST.
Each AST node comprises a node {\em type} representing some syntactic element, the {\em file range} ({\em source file} and {\em offset} range) from which the element was parsed, and zero or more children of sub-elements.
The AST is {\em unified} so that the shared language features are mapped to the same representation, as illustrated in \autoref{fig:uast}.
Although \verb+tree-sitter+ does not guarantee perfect parsing, this is tolerable under our approximate design.
Our approach also avoids heavyweight and specialized compiler frameworks necessary for precise parsing (\eg \verb+clang+~\cite{Chris04} for \verb+C+, \verb+javac+~\cite{Arnold05} for Java, and CPython~\cite{Python} for Python), significantly simplifying the implementation.

\myparagraph{CFG Construction}
The next step is to {\em lower} the AST into a {\em Control Flow Graph} (CFG), where each node represents a statement or a condition, followed by one (or more) control-flow edges to successor nodes.
The CFG is language agnostic, and common language features (\eg~{\em if-then-else}, {\em while-loops}, etc.) are lowered into common CFG patterns.

\myparagraph{Slice generation}
The path partitioning is generated using \autoref{alg:gen}.
For each partition $\Pi$, a corresponding truncated sub-program $T_\Pi$ is generated by substituting non-covered nodes/sub-programs with an \verb+assume(0)+ statement, followed by simplification.
From this, a truncated slice $S_\Pi$ is generated by applying Weiser's back-slicing algorithm (as illustrated in \autoref{alg:slice}).
Here, the variables from the post-condition ($\QQ$) are used as the slicing criterion, and \autoref{alg:slice} removes all nodes from $T_\Pi$ of which $\QQ$ is not control- or data-flow dependent.

\subsection{Truncated Slice Rendering}

So far, each slice ($S_\Pi$) is represented as a {\em sub-graph} of the (truncated) CFG.
Our ultimate target is an off-the-shelf LLM, meaning that the slice must be {\em rendered} back into a generic text form.
For this, we leverage the underlying AST representation based on {\em file ranges}.
For the rendering algorithm, we use an {\em interval tree} ($I$) mapping ranges to strings.
For each CFG node (and the corresponding AST statement $s$), we insert all $s \in S_\Pi$ into $I$.

\myparagraph{Context}
We also include the AST path from the root down to each sliced ($s$), such as the enclosing {\em function}, control-flow context ({\em if-then-else}, {\em while-loops}, etc.), or {\em class definitions} (for Java).
This ensures that the rendered slice is coherent (reparsable) code, rather than a collection of isolated statements.
Furthermore, it is often useful to include all relevant {\em non-executable} AST nodes from which each $s$ depends, including
{\em variable declarations}, {\em type declarations}, {\em member declarations}, {\em function declarations}, {\em global declarations}, {\em macro definitions}, etc.
Such dependencies are matched based on the AST name.
For example, given the executable statement (\verb+xs = insert(xs, x)+), then any function declaration or macro definition matching the name ``\verb+insert+'' can be included in the context.
Since we rely on lightweight parsing and not a compiler front-end with semantic information, name-based matching may over-approximate dependencies.
However, this is allowable under our approximate workflow, and the LLM will often ignore irrelevant context.

The slice renderer also preserves the original formatting, including the preceding and succeeding whitespace and comments.
Preserving such information is not strictly necessary for \verb+C+ and Java, but is essential for whitespace-sensitive languages such as Python.
Furthermore, source-code comments can provide additional context, such as the intent of programmers, which can assist LLM inference.

\myparagraph{Example}
An example rendered slice is illustrated in \autoref{fig:workflow} (d).
Here, all sliced executable statements are preserved, as well as the enclosing context (\eg function declaration for \verb+f()+).
Furthermore, the inner {\em if-then-else} has been simplified to an assertion, and the formatting (whitespace) has been preserved.
The resulting slice is coherent \verb+C+ code in its own right.
Crucially, each path through the slice corresponds to an equivalent path in the original programming, meaning that any inference on the slice is also a valid inference for the original code.

\myparagraph{LLM Inference}
Once the rendered slice has been generated, the final step is prompt construction and LLM inference.
This step can be highly customized,
but for our basic design, 
we use a prompt structure that mirrors a Hoare triple $\triple{\PP}{S}{\QQ}$ where $\PP$ is a pre-condition, $\QQ$ is a post-condition, and $S$ is the generated slice (see \autoref{ex:motive}, \autoref{ex:slice}, and \autoref{fig:workflow} (e)).
In addition, some basic instructions for the LLM are provided,
such as the output format.
Here, we assume that the LLM will generate one of two possible responses to the prompt, namely \verb+PASS+ (the post-condition holds) or \verb+FAIL+ (the post-condition does not hold).

\subsection{Implementation}
We have implemented our workflow in the form of the \tool tool.
\tool takes as input a program $C$ in a supported programming language (currently \verb+C+/Java/Python), pre- and post-conditions ($\PP$ and $\QQ$) expressed as code, constraints, or natural language.
\tool automatically decomposes the program into a sequence of truncated slices, and then constructs a sequence of prompts (\aka verification conditions) to be sent to the LLM for inference.
The slices are ordered by size to find the {\em least} counterexample to the post-condition where applicable.
\tool is also language-agnostic, except for the parser and some elements of the renderer.
Furthermore, \tool is lightweight and approximate by design---significantly simplifying the implementation (\ie it does not rely on language-specific compiler infrastructure).
This also reflects the nature of LLM inference, which is heuristic by nature rather than relying on precise parsing and semantic analysis, but can still make useful inferences for many real-world applications.
The implementation is also designed to be LLM-agnostic: since most modern LLMs expose a common API interface, \tool can be configured to query different models without changes to the core workflow.
\section{Evaluation}
To evaluate the effectiveness of LLM-based symbolic execution implemented by \tool, we answer the following research questions:

\begin{description}[leftmargin=*]
\item[\textbf{RQ.1}] \textbf{(Accuracy)} What is the accuracy of \tool compared to baselines?
\item[\textbf{RQ.2}] \textbf{(Scalability)} Can \tool scale to real-world large-scale programs?
\item[\textbf{RQ.3}] \textbf{(Language Agnosticism)} Can \tool support multiple programming languages?
\end{description}

\subsection{Experimental Setup}

\subsubsection{Dataset} We evaluated \tool on two widely-used datasets: \textsc{REval} and \textsc{CodeForces}. \textsc{REval}~\cite{Chen24} is designed to evaluate the code-reasoning abilities of LLMs. It consists of 85 Python solutions to LeetCode problems, some of which include accompanying test suites. The subjects in \textsc{REval} usually employ multiple control flow constructs, such as nested conditions and loops. This dataset was released in June 2024. In addition, we included the \textsc{CodeForces} dataset, which presents more challenging programming tasks. \textsc{CodeForces} is a programming contest platform that regularly publishes new problem sets. From this platform, we collected 662 subjects written in both Python and \verb+C+, published in June 2025. To the best of our knowledge, this is the most recent and largest open dataset currently available from the platform. The subjects in both datasets typically span around 100 lines of code, with token counts reaching up to 900. 

To conduct the evaluation, we also needed to construct non-trivial Hoare triples for each subject. This presented a challenge, as both \textsc{REval} and \textsc{CodeForces} provide only problem descriptions in natural language, source-code solutions, and test suites in some cases. To generate Hoare triples, we employed three different strategies:

\begin{itemize}[leftmargin=*]
\item \textsc{REval-Desc}: We treated the program descriptions in natural language as the post-conditions, and used any restrictions on test input when available as the pre-conditions. This strategy was applied to the \textsc{REval} dataset to construct \textsc{REval-Desc} with the corresponding pre- and post-conditions.

\item \textsc{CodeForces-Automatic}: In this strategy, we used an LLM to automatically generate pre- and post-conditions in natural language based on the program descriptions from \textsc{CodeForces}. This resulted in the dataset \textsc{CodeForces-Automatic} for evaluation.

\item \textsc{Mixed-Manual}: Here, we manually annotated formal pre- and post-conditions in executable code based on implicit properties of the original problems, e.g., by adding assertions to verify the correctness of a sorting algorithm. Given the substantial manual effort required, we randomly sampled 30 subjects in both \textsc{REval} and \textsc{CodeForces}, covering both Python and \verb|C| solutions, and applied this strategy to them. These formed another dataset \textsc{Mixed-Manual}. 

\end{itemize}

\subsubsection{Baselines} For comparison, we selected two main types of baselines. For the first baseline, we used an ``ad-hoc'' LLM-based program analysis over the entire subject without decomposition. 
This baseline queries the LLM for a counterexample (if one exists) or whether the post-condition always holds for inputs that satisfy the pre-condition. 
This baseline aims to show the effectiveness of the systematic partitioning provided by our tool \tool.

The second baseline includes traditional symbolic execution tools.
For Python subjects, we selected CrossHair~\cite{CrossHair} as a baseline. Although there are other traditional symbolic execution tools for Python, such as PyExZ3~\cite{PyExZ3} and pySym~\cite{pySym}, they are no longer maintained (unmaintained for six years or more at the time of writing). For the \verb+C+ subjects, we compared our tool with KLEE~\cite{Cadar08}, a well-established and mature traditional symbolic execution tool. A practical issue of these traditional tools is the lack of expressivity in the pre- and post-conditions, which must be specified in code or other formal languages. Therefore, we only compare \tool with traditional tools on the \textsc{Mixed-Manual} dataset, which contains conditions in the formal form of executable code.

\subsubsection{Test Models}
We selected five of the most popular openly available language models (at the time of writing)\footnote{\url{https://ollama.com/search}}. Where possible, we also included different numbers of parameters to demonstrate the potential impact of the parameter number. 

\begin{itemize}[leftmargin=*]
\item Meta's \textsc{Llama3} series~\cite{Llama3}: Instruction-tuned model optimized for general use cases such as dialogue and chatting, which was released in December 2024. 

\item Microsoft's \textsc{Phi-4}~\cite{Phi-4}: Model trained on high-quality data from filtered public domain websites and acquired academic books, designed to aid research on language models, which was released in December 2024.

\item DeepSeek's \textsc{DeepSeek-R1}~\cite{DeepSeek-R1}: The first-generation reasoning model from DeepSeek, achieving performance comparable to OpenAI-o1 across math, code, and reasoning tasks with much smaller parameters, which was released in January 2025.

\item Google's \textsc{Gemma3}~\cite{Gemma3}: A lightweight family of multimodal models that features a larger context window of 128K, and released in March 2025. 

\item Alibaba's \textsc{Qwen3}~\cite{Qwen3}: The latest version of the \textsc{Qwen} model family, designed to advance performance, efficiency, and multilingual capabilities, and it was released on April 2025.
\end{itemize}

All experiments were carried out using \verb+ollama+ version v0.9.6. Each experiment was repeated three times, and we report the average results rounded to the nearest integer. 

\subsection{RQ.1: Accuracy} \label{sec:rq1}

\subsubsection*{Method} 
To evaluate the accuracy of \tool in generating counterexamples and verifying post-conditions, we compared \tool with the ``ad-hoc'' LLM-based program analysis on \textsc{REval-Desc}, \textsc{CodeForces-Automatic}, and \textsc{Mixed-Manual}. In addition, we compared \tool with traditional symbolic execution tools CrossHair and KLEE on \textsc{Mixed-Manual} with formal pre- and post-conditions expressed in executable Python or \verb+C+ code. We did not conduct the comparison on other datasets, since traditional tools cannot handle post-conditions in natural language. In each subject, we executed \tool and baselines, and reported their results. The accuracy in each subject was measured by comparing the result with the ground truth derived from the provided test suites, if available, or manual verification.

\subsubsection*{Main result}

\autoref{fig:RQ1-results} shows the accuracy of \tool compared to the ``ad-hoc'' LLM-based program analysis across different datasets under different models. The experiments were conducted under the list of models (\textit{Model}). For each dataset, we report the total number of subjects analyzed (\textit{Total}), how many subjects were correctly analyzed by the tool (\textit{Correct}), and the corresponding accuracy rate (\textit{Accuracy}). Overall, \tool consistently achieves a higher accuracy than the baseline in all datasets using different forms of Hoare triples. On average, \tool improves accuracy from 84.7\% to 90.6\% on \textsc{REval-Desc}, from 79.2\% to 86.4\% on \textsc{CodeForces-Automatic}, and from 66.7\% to 73.3\% on \textsc{Mixed-Manual}. Notably, on these three (3) datasets under eight (8) language models, \tool outperforms the baseline in 20 out of 24 experiments (\ie 83.3\% scenarios), with the only exceptions on \textsc{REval-Desc} and \textsc{Mixed-Manual} when using the model \textsc{Gemma3-27B}. These results demonstrate the effectiveness of our approach in improving the accuracy of program analysis. 

\begin{table}[H]
{
\small
\caption{Accuracy of \tool and the ``ad-hoc'' LLM-based program analysis under different datasets.\label{fig:RQ1-results}}
\setlength\tabcolsep{1.7pt}
\begin{tabular}{c|c|rrr|rrr|rrr}
\toprule
\multirow{2}{*}{Model} & \multirow{2}{*}{Method} & \multicolumn{3}{c|}{\textsc{REval-Desc}} & \multicolumn{3}{c|}{\rev{\textsc{CodeForces-Automatic}}} & \multicolumn{3}{c}{\textsc{Mixed-Manual}} \\
\cline{3-11}
& & Total & Correct & Accuracy & \rev{Total} & \rev{Correct} & \rev{Accuracy} & Total & Correct & Accuracy \\
\hline
\hline
\multirow{2}{*}{\textsc{Llama3-8B}} & \tool & 85 & 82 & {\bf 96.5\%} & \rev{662} & \rev{654} & \rev{\bf 98.8\%} & 30 & 23 & {\bf 76.7\%} \\
% \cline{2-11}
& Baseline & 85 & 79 & 92.9\% & \rev{662} & \rev{653} & \rev{98.6\%} & 30 & 22 & 73.3\% \\
\hline
\multirow{2}{*}{\textsc{Llama3.1-8B}} & \tool & 85 & 84 & {\bf 98.8\%} & \rev{662} & \rev{616} & \rev{\bf 93.1\%} & 30 & 23 & {\bf 76.7\%} \\
% \cline{2-11}
& Baseline & 85 & 71 & 83.5\% & \rev{662} & \rev{547} & \rev{82.6\%} & 30 & 18 & 60.0\% \\
\hline
\multirow{2}{*}{\textsc{Llama3.3-70B}} & \tool & 85 & 72 & {\bf 84.7\%} & \rev{662} & \rev{617} & \rev{\bf 93.2\%} & 30 & 26 & {\bf 86.7\%} \\
% \cline{2-11}
& Baseline & 85 & 70 & 82.4\% & \rev{662} & \rev{589} & \rev{89.0\%} & 30 & 21 & 70.0\% \\
\hline
\multirow{2}{*}{\textsc{Phi-4-14B}} & \tool & 85 & 74 & {\bf 87.1\%} & \rev{662} & \rev{626} & \rev{\bf 94.6\%} & 30 & 24 & 80.0\% \\
% \cline{2-11}
& Baseline & 85 & 72 & 84.7\% & \rev{662} & \rev{584} & \rev{88.2\%} & 30 & 24 & 80.0\% \\
\hline
\multirow{2}{*}{\textsc{DeepSeek-R1-70B}} & \tool & 85 & 85 & {\bf 100.0\%} & \rev{662} & \rev{561} & \rev{\bf 84.7\%} & 30 & 19 & {\bf 63.3\%} \\
% \cline{2-11}
& Baseline & 85 & 82 & 96.5\% & \rev{662} & \rev{508} & \rev{76.7\%} & 30 & 17 & 56.7\% \\
\hline
\multirow{2}{*}{\textsc{Gemma3-4B}} & \tool & 85 & 73 & {\bf 85.9\%} & \rev{662} & \rev{305} & \rev{\bf 46.1\%} & 30 & 20 & {\bf 66.7\%} \\
% \cline{2-11}
& Baseline & 85 & 64 & 75.3\% & \rev{662} & \rev{265} & \rev{40.0\%} & 30 & 16 & 53.3\% \\
\hline
\multirow{2}{*}{\textsc{Gemma3-27B}} & \tool & 85 & 72 & 84.7\% & \rev{662} & \rev{598} & \rev{\bf 90.3\%} & 30 & 20 & 66.7\% \\
% \cline{2-11}
& Baseline & 85 & 76 & {\bf 89.4\%} & \rev{662} & \rev{547} & \rev{82.6\%} & 30 & 23 & {\bf 76.7\%} \\
\hline
\multirow{2}{*}{\rev{\textsc{Qwen3-32B}}} & \rev{\tool} & \rev{85} & \rev{71} & \rev{\bf 83.5\%} & \rev{662} & \rev{601} & \rev{\bf 90.8\%} & \rev{30} & \rev{21} & \rev{70.0\%} \\
% \cline{2-11}
& \rev{Baseline} & \rev{85} & \rev{61} & \rev{71.8\%} & \rev{662} & \rev{499} & \rev{75.4\%} & \rev{30} & \rev{21} & \rev{70.0\%} \\
\hline
\hline
\multirow{2}{*}{Average} & \tool & 85 &  77 & {\bf 90.6\%} & \rev{662} & \rev{572} & \rev{\bf 86.4\%} & 30 & {22} & {\bf 73.3\%} \\
% \cline{2-11}
& Baseline & 85 & 72 & 84.7\% & \rev{662} & \rev{524} & \rev{79.2\%} & 30 & 20 & 66.7\% \\
\bottomrule
\end{tabular}
}
\setlength{\fboxsep}{0pt}
\end{table}

\setlength{\oldcolumnsep}{\columnsep}
\setlength{\columnsep}{2em}
\begin{wraptable}{r}{0.35\textwidth}
{
\small
\caption{Accuracy of \tool and traditional symbolic execution tools.}
\label{fig:RQ1-traditional-results}
\setlength\tabcolsep{3pt}
\hspace{-0.5em}
\begin{tabular}{c|rrr}
\toprule
 \multirow{2}{*}{Method} & \multicolumn{3}{c}{\textsc{Mixed-Manual}} \\
\cline{2-4}
 & Total & Correct & Accuracy \\
\hline
\hline
\tool & 30 & 22 & {\bf 73.3\%} \\
Baseline & 30 & 14 & 46.7\% \\
\bottomrule
\end{tabular}
}
\setlength{\fboxsep}{0pt}
\end{wraptable}

The comparison results with traditional symbolic execution tools are presented in \autoref{fig:RQ1-traditional-results}, which shows the accuracy of \tool, and CrossHair and KLEE on \textsc{Mixed-Manual} with formal post-conditions in the form of executable code. We can see that \tool significantly outperforms these traditional tools. \tool successfully analyzes 73.3\% of subjects, compared to just 46.7\% for traditional tools. This performance gap is due to the inherent limitations of traditional symbolic execution (as illustrated in \autoref{sec:se_limits}), such as the difficulties in reasoning about nested loops and complex constructs. Moreover, compared to traditional tools, \tool also demonstrates greater applicability by supporting expressive post-conditions, including those written in natural language.

\result{\tool improves accuracy over the baselines across different datasets. Moreover, \tool is more applicable than traditional symbolic execution tools, which can handle pre- and post-conditions expressed in either code or natural language.}

\setlength{\columnsep}{\oldcolumnsep}

\subsubsection*{Impact of different LLMs}

We evaluated \tool on multiple different LLMs, even with different numbers of parameters as shown in \autoref{fig:RQ1-results}: Meta's \textsc{Llama3}, \textsc{Llama3.1} and \textsc{Llama3.3} models with 8B and 70B parameters, Microsoft's 14B \textsc{Phi-4} model, the 70B distilled version of \textsc{DeepSeek-R1} model, Google's \textsc{Gemma3} model with 4B and 27B parameters, and Alibaba's 32B \textsc{Qwen3}. 

It is interesting to note that the accuracy varies significantly under different models, even with different model sizes. Among the models used, the most significant accuracy improvement stems from \textsc{Llama3.1-8B}, where \tool achieves 98.8\% accuracy compared to the baseline's 83.5\%. This highlights the benefits of partitioning in improving the accuracy of LLM reasoning for comparatively small models. In contrast, on larger models such as \textsc{Llama3.3-70B} and \textsc{DeepSeek-R1-70B}, such improvements in accuracy are more subtle, mainly due to the fact that these larger models inherently have a stronger reasoning ability. However, the result is consistently positive for larger models, with the exception of \textsc{Gemma3-27B}. This is a positive result---smaller LLMs can be run locally on consumer hardware, and \tool exhibits a clear benefit for these use cases. 

Under \textsc{Gemma3-27B}, \tool outperforms the baseline on \textsc{CodeForces-Automatic}, but shows weaknesses on \textsc{REval-Desc} and \textsc{Mixed-Manual}. Upon closer investigation, we attribute this to Gemma3-27B's tendency to overinterpret problem specifications and penalize even minor mismatches. For example, during slicing, \tool may omit certain input-handling statements if they are irrelevant to the target context, resulting in incomplete input-processing logic. Gemma3-32B penalizes such incompleteness and reports unsatisfactory when inputs are not fully processed, while Gemma3-4B tends to overlook these issues and focuses on reasoning about the main task.

\result{
\tool improves accuracy over the evaluated LLMs, where smaller LLMs benefit the most.
}

\subsubsection*{Prompt size}

\autoref{fig:RQ1-TC} shows the average token counts used in the LLM prompts for both the baseline and \tool across different datasets. To calculate the token count in each LLM prompt, we use the tool \verb|tiktoken|~\cite{tiktoken} with the corresponding Byte-Pair Encoding (BPE) token encoding, which is also used by GPT-4o mini~\cite{GPT4o-mini}. To clearly demonstrate the reduction ability of \tool, we also reported the reduction ratio achieved by \tool, compared with the baseline. 

As shown in \autoref{fig:RQ1-TC}, \tool effectively reduces token counts---reducing the length and complexity of the prompt---with average reductions of approximately 9.5\%, 25.3\%, and 26.3\% compared to the baseline for \textsc{REval-Desc}, \textsc{CodeForces-Automatic}, and \textsc{Mixed-Manual}, respectively.

\begin{table}[H]
{
\small
\caption{Average LLM token counts for the baseline and \tool under different datasets, along with the token reduction ratio achieved by \tool. Each value is reported in the form of \textsc{Mean} $\pm$ \textsc{Standard Deviation}. \label{fig:RQ1-TC}}
\setlength{\fboxsep}{0pt}
\begin{tabular}{c|c|c|c}
\toprule
Method & \textsc{REval-Desc} & \textsc{CodeForces-Automatic} & \textsc{Mixed-Manual} \\
\hline
\hline
Baseline & $224.8\pm85.9$ & $346.4\pm116.4$ & $146.9\pm48.3$ \\
\tool & $203.4\pm75.5$ & $258.6\pm102.1$ & $108.3\pm40.2$ \\
\hline
Reduction Ratio & 9.5\% & 25.3\% & 26.3\% \\
\bottomrule
\end{tabular}
}
\end{table}

\result{
\tool significantly reduces the average token count required for LLM queries to find the minimum counterexample across all tested datasets.
}

\subsubsection*{Case study}
To illustrate the advantages of \tool over the baselines, we use a concrete example in \autoref{fig:case-study}. This example consists of a buggy Python function shown in \autoref{fig:case-study}~(a), which is extracted from the \textsc{REval} dataset~\cite{Chen24}. Here, the pre- and post-conditions are formally annotated with the corresponding \verb|PRE| and \verb|POST| comments. The pre-condition specifies that $\texttt{value}$ is not empty, which is indicated by the (\verb|assume len(value) > 0|) statement located at the beginning of the function. The post-condition of the function states that $|\texttt{res}| \leq |\mathit{float}(\texttt{value})|$. This condition does not always hold, since the given function rounds up any decimal input ending in ``\verb|.5|''. For example, $\mathit{closest\_integer}(\texttt{'1.5'}) = 2$, which violates the post-condition.

We find that most LLMs cannot correctly reason over the whole program. A typical response from \textsc{Llama3.1-8B}~\cite{Llama3} when fed with the entire original program is as follows:
\begin{quote}
``{\em For inputs with a single decimal point, removing trailing zeros ensures that the absolute value of $\texttt{res}$ (the rounded integer) is less than or equal to the absolute value of the original float ($\mathit{float}(\texttt{value})$).}''
\end{quote}
The LLM then proceeds to incorrectly declare that the post-condition of the program will always be satisfied. In this case, the LLM is distracted over the trailing zero removal branch so much, thereby it does not actually try to analyze what the program is supposed to do with the converted value---\ie an example of hallucination.

\begin{figure}
    \centering
    \begin{minipage}{0.47\textwidth}
        \centering
        {\footnotesize \textcolor{red!60!black}{(a)} original Python function}
        \begin{lstlisting}[language=Python, morekeywords={assert, assume}]
def closest_integer(value):
    assume len(value) > 0  # PRE
    
    if value.count('.') == 1:
        # remove trailing zeros
        while value[-1] == '0':
            value = value[:-1]
    
    num = float(value)
    if value[-2:] == '.5':
        if num > 0:
            res = ceil(num)
        else:
            res = floor(num)
    elif len(value) > 1 or value[0] != '0':
        res = int(round(num))
    else:
        res = 0
    
    assert abs(res) <= abs(float(value))  # POST
	\end{lstlisting}
    \end{minipage}
    \hspace{0.2cm}
    \begin{minipage}{0.45\textwidth}
        \centering
        {\footnotesize \textcolor{red!60!black}{(b)} truncated slice \#1}
        \begin{lstlisting}[mathescape, language=Python, morekeywords={assert, assume}]
def closest_integer(value):
    assume len(value) > 0  # PRE
    if value.count('.') == 1:
        # remove trailing zeros
        while value[-1] == '0':
            value = value[:-1]
    num = float(value)
    assume value[-2:] != '.5'
    assume len(value) > 1 or value[0] != '0'
    res = int(round(num))        
    assert abs(res) <= abs(float(value))  # POST
        \end{lstlisting}
        
        \vspace{0.2cm}
        {\footnotesize \textcolor{red!60!black}{(c)} truncated slice \#2}
        \begin{lstlisting}[mathescape, language=Python, morekeywords={assert, assume}]
def closest_integer(value):
    assume len(value) > 0  # PRE
    assume value.count('.') == 1 
    assume value[-1] != '0'
    num = float(value)
    assume value[-2:] == '.5'
    assume num > 0
    res = ceil(num)   
    assert abs(res) <= abs(float(value))  # POST
        \end{lstlisting}
    \end{minipage}
\caption{\textcolor{red!60!black}{(a)} is a buggy Python program that implements a simple function to take a value (string) representing a number and return the closest integer to it. In cases where the number is equidistant from two integers, it rounds away from zero. This example includes an unbounded loop (\texttt{while value[-1] == '0': ...}), invokes third-party library APIs (\texttt{ceil}, \texttt{floor}, and \texttt{abs}), and introduces complex language constructs like list slicing. \textcolor{red!60!black}{(b)} and \textcolor{red!60!black}{(c)} show the corresponding program slides for inputs \texttt{"0.0"} and \texttt{"2.5"}, respectively.}
\label{fig:case-study}
\end{figure}

In contrast, \tool decomposes the input program into truncated slices. Two of the generated slices from the original Python function are shown in \autoref{fig:case-study}~(b) and (c). As illustrated in this example, we can see that \tool significantly reduces the complexity and size of the resulting truncated slices. In \autoref{fig:case-study}~(b), only the \verb|elif| branch is considered, with the others truncated, forming implicit pre-conditions in the form of \verb|assume|-statements.
In \autoref{fig:case-study}~(c), only the second \verb+if+ branch is assumed to be taken. In addition, the original Python code (a) contains 430 tokens, which is reduced to 103 tokens (${\sim}76\%$ reduction) for (b) and 69 tokens (${\sim}84\%$ reduction) for (c).
Together with the reduced CFG complexity, the slices form a smaller and more targeted prompt, increasing the chance of a correct validation result.
Using these truncated slices, the LLM correctly infers that the post-condition holds for slice (b), but does not hold for slice (c).
For example, \textsc{Llama3.1-8B} outputs the following for (c) (emphasis original):
\begin{quote}
``{In other words, given an input that satisfies the precondition (but has a decimal point followed by \verb+'.5'+), the postcondition will not be satisfied. The result of \verb+ceil+ will exceed the absolute value of the original float.
Therefore, we conclude that the postcondition indicated by the \verb+assert+ statement with a \verb_POST_ comment is {\bf not always satisfiable}.}''
\end{quote}
Since there exists a counterexample for (c), \tool has shown that the post-condition does not hold for the original program (a).

We also attempted to analyze \autoref{fig:case-study}~(a) using CrossHair, a traditional symbolic execution engine for Python. However, CrossHair terminates after failing to find any violation of the post-condition. We believe that this is due to an incomplete search or incomplete solving of the path constraint by the underlying SMT solver (z3). In contrast, \tool is capable of exhaustively enumerating all path partitions, and the underlying LLM correctly infers the corresponding slice. This case study, although it involves a relatively simple program, highlights the limitations of traditional methods when compared to \tool.

\subsection{RQ.2: Scalability}

\subsubsection*{Method} To evaluate the scalability of \tool, we selected eight (8) real-world large-scale programs: X11, ncurses, nano, cURL, SQLite3, OpenSSH, FFmpeg, and Zstd, which are widely-used benchmarks for evaluating the performance of bug-finding techniques. \autoref{fig:large_subjects} summarizes the statistics of these subject programs, including the lines of code (\emph{\#LoC}) and the token counts (\emph{\#Token}). These subjects have lines of code up to 250K and have 197K--4960K token counts. In this experiment, we examined whether \tool could successfully analyze the bugs recently found by existing bug detectors~\cite{Meng24}. These bugs were found in several X11 client applications (xinput, xlsclients, xmodmap, xset, xwininfo), as well as in ncurses, SQLite3, and other common UNIX libraries, respectively. To construct post-conditions, we negated the crash condition, such as ($\mathit{ptr}~\texttt{!= NULL}$) for \texttt{NULL}-pointer dereference. To demonstrate the effectiveness of \tool's decompositions, we evaluated three forms: only relevant source files included (\emph{File}), only relevant functions included (\emph{Func}), and relevant code sliced by \tool (\emph{Slice}) along a path that exercises the bug.

\begin{table}
\small
\caption{Statistics of large-scale programs used, including lines of code (\#LoC) and total token counts (\#Token).}
\label{fig:large_subjects}
\begin{tabular}{l|r|r|r|r|r|r|r|r}
\toprule
Subject & X11   & \rev{ncurses} & \rev{nano} & \rev{cURL}  & \rev{SQLite3} & \rev{OpenSSH} & \rev{FFmpeg} & \rev{Zstd} \\
\hline
\hline
\#LoC   & 244K  & \rev{54K}     & \rev{24K}  & \rev{172K}  & \rev{225K}    & \rev{144K}    & \rev{255K}   & \rev{35K} \\
\hline
\#Token & 4960K & \rev{438K}    & \rev{197K} & \rev{1379K} & \rev{2341K}   & \rev{1427K}   & \rev{2761K}  & \rev{370K} \\
\bottomrule
\end{tabular}
\end{table}

\begin{table}
{
\small
\caption{The token counts for file-based (File), function-based (Func), and slice-based (Slice) decompositions of real-world bugs from X11 client applications and other examples.
Here ({\cmark}) means the LLM correct detects the bug, ({\xmark}) means the bug was not detected, and ({$\boldsymbol{-}$}) means the token limit for the model was exceeded.
\label{fig:scale}}
\setlength\tabcolsep{4pt}
\begin{tabular}{c|c|rrr|ccc|ccc}
\toprule
\multirow{2}{*}{Subject} & \multirow{2}{*}{Bug} & \multicolumn{3}{c|}{Token Count} & \multicolumn{3}{c|}{\textsc{Llama3.1-8B}} & \multicolumn{3}{c}{\textsc{Gemma3-4B}} \\
\cline{3-11}
 &  & File & Func & Slice & File & Func & Slice & File & Func & Slice \\
\hline
\hline
\texttt{xinput} & {\em \texttt{NULL}-pointer} & 4281 & 1197 & 178 &
\xmark & \xmark & \cmark & \xmark & \xmark & \cmark \\
\hline
\texttt{xlsclients} & {\em \texttt{NULL}-pointer} & 140120 & 1090 & 229 &
$\boldsymbol{-}$ & \xmark & \xmark & $\boldsymbol{-}$ & \xmark & \xmark \\
\hline
\texttt{xmodmap} & {\em \texttt{NULL}-pointer} & 18683 & 2189 & 279 &
\xmark & \xmark & \cmark & \xmark & \xmark & \cmark \\
\hline
\texttt{xset} & {\em Divide-by-zero} & 17078 & 3164 & 506 &
\xmark & \cmark & \cmark & \xmark & \cmark & \cmark \\
\hline
\texttt{xwininfo} & {\em Buffer-overflow} & 152020 & 2560 & 597 &
$\boldsymbol{-}$ & \xmark & \xmark & $\boldsymbol{-}$ & \xmark & \xmark \\
\hline
% \hline
\texttt{\rev{ncurses}} & \rev{\em \texttt{NULL}-pointer} & \rev{1446} & \rev{271} & \rev{201} &
\cmark & \cmark & \cmark & \xmark & \cmark & \cmark \\
\hline
% \hline
\texttt{\rev{nano}} & \rev{\em Environment race} & \rev{6716} & \rev{698} & \rev{125} &
\xmark & \cmark & \cmark & \xmark & \xmark & \cmark \\
\hline
% \hline
\texttt{\rev{cURL}} & \rev{\em \texttt{assert}-trigger} & \rev{21285} & \rev{5581} & \rev{202} &
\xmark & \cmark & \cmark & \xmark & \xmark & \cmark \\
\hline
% \hline
\texttt{\rev{SQLite3}} & \rev{\em \texttt{NULL}-pointer} & \rev{2341085} & \rev{213088} & \rev{4691} &
$\boldsymbol{-}$ & $\boldsymbol{-}$ & \xmark & $\boldsymbol{-}$ & $\boldsymbol{-}$ & \xmark \\
\hline
% \hline
\texttt{\rev{OpenSSH}} & \rev{\em \texttt{assert}-trigger} & \rev{10164} & \rev{446} & \rev{208} &
\xmark & \cmark & \cmark & \xmark & \cmark & \cmark \\
\hline
% \hline
\texttt{\rev{FFmpeg}} & \rev{\em Buffer-overflow} & \rev{118456} & \rev{19242} & \rev{4370} &
\xmark & \xmark & \cmark & \xmark & \xmark & \cmark \\
\hline
% \hline
\texttt{\rev{Zstd}} & \rev{\em \texttt{NULL}-pointer} & \rev{60813} & \rev{936} & \rev{328} &
\xmark & \xmark & \cmark & \xmark & \xmark & \cmark \\
\hline
\hline
Accuracy & --- & --- & --- & --- & 8.3\% & 41.7\% & {\bf 75.0\%} & 0\% & 25.0\% & {\bf 75.0\%} \\
\bottomrule
\end{tabular}
}
\end{table}

\subsubsection*{Results} The results are shown in \autoref{fig:scale}. \tool demonstrates strong scalability to real-world, large-scale programs, with token counts for all cases remaining within the context window of LLMs, making LLM-based symbolic execution feasible. Moreover, \tool achieves high accuracy, successfully analyzing 75.0\% of the bugs. In contrast, traditional symbolic execution tools suffer from the well-known path-exploration problem and fail to scale to these programs. Additionally, the token counts for these subjects are too large for the ad-hoc LLM-based analysis within a single prompt, making decomposition necessary. 

\setlength{\oldcolumnsep}{\columnsep}
\setlength{\columnsep}{2em}
\begin{wrapfigure}{r}{0.5\textwidth}{}
    \small
    \hspace{-1cm}
    \begin{lstlisting}[mathescape, language=C, morekeywords={assert, assume}, xrightmargin=1em]
XIDeviceInfo* XIQueryDevice(Display *dpy, 
              int deviceid, int *ndevices_return)
{
  xXIQueryDeviceReply reply;
  XExtDisplayInfo *extinfo = XInput_find_display(dpy);
  $...$
  *ndevices_return = -1;
  return NULL;
}
void XIFreeDeviceInfo(XIDeviceInfo* info)
{
  // POST: info != NULL
}
static int list_xi2(Display *display,
         enum print_format format)
{
  // PRE: true
  int ndev;
  XIDeviceInfo *info, *dev;
  info = XIQueryDevice(display, XIAllDevices, &ndev);
  XIFreeDeviceInfo(info);
}
\end{lstlisting}
\caption{Relevant code sliced for the bug in xinput.}
\label{fig:xinput}
\end{wrapfigure}

Compared to na\"ive decomposition methods based on files and functions, \tool's slice-based decomposition method is significantly more effective. Using slicing, \tool successfully analyzes 75.0\% of the bugs, while the file-based and function-based methods achieve only 8.3\% and 41.7\% success rates, respectively. To further show the effectiveness of \tool's slicing method, we examine the bug in xinput. The bug resides in a program that contains 244K sLOC (${\sim}$4.96M tokens), but the sliced code is reduced to just 178 tokens, as shown in \autoref{fig:xinput}. With the code size and complexity dramatically reduced, the bug becomes apparent: the \verb+XIQueryDevice()+ function returns \verb+NULL+ (derived from an error-handling path), and this value is immediately passed to \verb+XIFreeDeviceInfo()+ (the intermediate code was removed by the slicer), thus violating the post-condition. After truncation and slicing, the violation is readily apparent even to small LLMs. In contrast, the other decompositions based on files and functions result in 4,281 and 1,197 tokens, respectively. They include large amounts of irrelevant information that obscure the bug and cause LLMs to miss it.   

\result{\tool scales to real-world large-scale programs with 75\% accuracy. Truncated slicing significantly reduces prompt size and complexity---improving the accuracy of LLM inference.}

\setlength{\columnsep}{\oldcolumnsep}

\subsection{RQ.3: Language Agnosticism}

\subsubsection*{Method} \tool implements a lightweight workflow that is language-agnostic. To evaluate \tool's ability to analyze programs written in different programming languages, we conducted this experiment using the same set of subjects implemented in Python, \verb+C+, and Java, respectively. Specifically, we use GPT-4o-mini to automatically translate the \textsc{REval-Desc} dataset (originally written in Python) into equivalent \verb+C+ and Java versions. These translated programs form the \textsc{C-REval-Desc} and \textsc{Java-REval-Desc} datasets, each consisting of 85 programs. We evaluated the three datasets using the same LLMs and framework as in RQ.1, and then reported the average results. 

\begin{table}
{
\small
\caption{Accuracy of \tool and the baseline on the Python, C, and Java versions of \textsc{Reval-Desc}.}
\label{fig:RQ3-results}
\setlength\tabcolsep{3.5pt}
\begin{tabular}{c|c|rrr|rrr|rrr}
\toprule
\multirow{2}{*}{Model} & \multirow{2}{*}{Method} & \multicolumn{3}{c|}{\textsc{Python-REval-Desc}} & \multicolumn{3}{c|}{\textsc{C-REval-Desc}} & \multicolumn{3}{c}{\textsc{Java-REval-Desc}} \\
\cline{3-11}
& & Total & Correct & Accuracy & Total & Correct & Accuracy & Total & Correct & Accuracy \\
\hline
\hline
\multirow{2}{*}{Average} & \tool & 85 & 77 & {\bf 90.6\%} & 85 & 74 & {\bf 87.1\%} & 85 & 76 & {\bf 89.4\%} \\
& Baseline & 85 & 72 & 84.7\% & 85 & 67 & 78.8\% & 85 & 71 & 83.5\% \\
\bottomrule
\end{tabular}
}
\setlength{\fboxsep}{0pt}
\end{table}

\subsubsection*{Results} The results are illustrated in \autoref{fig:RQ3-results}. Across these three datasets implemented in different programming languages, \tool maintains similar performance and consistently improves the accuracy of the analysis. Specifically, compared to the baseline---the ad-hoc LLM-based analysis, \tool increases the accuracy of the analysis from 84.7\% to 90.6\% in Python, 78.8\% to 87.1\% in \verb|C|, and 83.5\% to 89.4\% in Java, respectively. These results demonstrate the versatility of \tool in handling multi-language programs, which is the direct result of its language-agnostic workflow and the inherent ability of LLM to understand multiple languages. Notably, these improvements are achieved without reliance on precise parsers and/or language-specific compiler front-ends.

To evaluate traditional symbolic execution tools such as CrossHair and KLEE, annotating formal post-conditions requires significant manual effort. In RQ.1 (\autoref{sec:rq1}), we manually annotated formal post-conditions and have shown the effectiveness of \tool in the programs with formal post-conditions compared to traditional symbolic execution. In this experiment, we explored the potential of using LLMs to automatically generate post-conditions based on specifications and translate them into code suitable for symbolic execution. Specifically, for CrossHair, we instructed LLM to generate the PEP 315-suitable contract specification, while we prompted LLM to generate suitable calls to \verb+klee_make_symbolic+ for KLEE. However, the results were far from satisfactory. The LLMs consistently failed to translate natural language post-condition specifications into concrete, executable code. The generated code frequently contains compiler or runtime errors, including incomplete implementations, missing function bodies, or absent header and import statements. For example, CrossHair was only able to analyze fewer than 10\% of the translated dataset (8 out of 85 subjects). These findings indicate that substantial manual effort is still necessary when using traditional symbolic execution tools.

\result{Across the same set of subjects implemented in different programming languages (\texttt{C}, Python, and Java), \tool maintains similar performance and achieves consistent improvements in the accuracy of the analysis.}

\subsection{Discussion}

Our results show that path-based decomposition of program analysis tasks is effective at improving LLM-based program analysis, especially for small LLMs and real-world problems.
The significance is that it allows for higher accuracy to be achieved with smaller models that can be run on consumer-grade hardware.
In addition, we prove that a lightweight and language agnostic workflow is feasible and can still achieve good results.

\subsubsection*{Limitations}
LLM-based program analysis is applicable to program analysis tasks that cannot be handled by traditional means.
That said, the approximate reasoning of LLMs is not suitable for all applications, even with improved accuracy.
LLMs are also unlikely to ever achieve the same accuracy as traditional solvers for some tasks, such as solving systems of linear equations.
As such, we propose LLM-based symbolic execution as a complementary method that does not necessarily replace traditional approaches for all use cases.
Another limitation is that path-based decomposition may still explode, even when our approach is guaranteed to terminate.
This is an inherent limitation of path-based reasoning.
However, we believe that the decomposition based on the truncated slice is a significant mitigation.

\subsubsection*{Threat to validity}
A primary concern is the potential for data contamination in LLMs~\cite{Sallou24}, where our evaluation datasets may have been included in the models' training data. To mitigate this issue, we selected the subjects from \textsc{CodeForces} that were released in June 2025, after the models' pre-training cutoff date. As shown in \autoref{fig:RQ1-results}, \tool consistently outperforms the baseline on these subjects. This suggests that the performance of \tool is not a result of data contamination.
\section{Related Work}

\subsubsection*{Static program analysis via symbolic execution}
Symbolic execution~\cite{King76} is an established method for static programs whose origins also relate to early work formalizing programs as mathematical logic.
The idea is to execute symbolic states, representing sets of concrete inputs, allowing for the exhaustive exploration of program behavior.
Over the decades, many different symbolic execution engines and frameworks have been developed, such as:
KLEE~\cite{Cadar08},
Owi~\cite{Andr24},
{\em Symbolic PathFinder} (SPF)~\cite{Spf10},
{\em Java PathFinder} (JPF)~\cite{Visser04},
CrossHair~\cite{CrossHair},
Angr~\cite{Angr},
S2E~\cite{Chipounov12},
PyExZ3~\cite{PyExZ3}, etc.
Unlike our approach, such traditional tools translate paths into some underlying formal language for theorem proving, thus inheriting many of the limitations discussed in this paper.
Furthermore, most existing tools are specialized to a specific language (\verb+C+, Java, binary, etc.) and are closely integrated into specific compiler frameworks (\eg LLVM~\cite{Chris04} for KLEE).
That said, LLMs use a fundamentally different type of reasoning compared to the deductive reasoning of theorem provers.
As such, traditional approaches are suitable for problems that can be handled by traditional methods and for applications where perfect accuracy is required.

\subsubsection*{LLM-based program analysis}
One recent alternative to traditional program analysis methods is {\em Large Language Models} (LLMs).
LLMs are very general tools and can be applied to a wide variety of tasks, including {\em fuzzing}~\cite{Asmita24, chatafl}, {\em vulnerability detection}~\cite{Zhou24}, and {\em program repair}~\cite{Fan23}.
Another recent innovation is LLM-based {\em agents}~\cite{Xi25}, which are algorithms where decisions are made by the LLM.
Our core \autoref{alg:symexe} is traditional and not agent-based.
However, an agentized version could be made, but decisions (\eg which branch to explore first) could be deferred to the LLM.

\subsubsection*{Intersection between symbolic execution and LLMs}
There are some other nascent works that combine LLMs and symbolic execution. HyLLfuzz~\cite{hyllfuzz} focuses on improving concolic execution integrated in hybrid fuzzing.
LLM-Sym~\cite{wang24} is an agent-based symbolic execution framework for Python code that uses an LLM to translate paths into traditional {\em path constraints} suitable for solving via z3~\cite{DeMoura08}.
LLM-Sym is fundamentally different in that our approach avoids translation altogether, instead directly using the LLM itself as a solver.
Since LLM-Sym still uses translation to z3, it inherits many of the limitations of traditional symbolic execution engines discussed in this paper.
Similarly, Loopy~\cite{kamath23} aims to discover {\em loop invariants} using LLMs, which can then be applied to symbolic analysis.
Our approach avoids the need for invariant discovery, since it is not based on translation.

\section{Conclusion}

In this paper, we introduce a variant of symbolic execution that uses an LLM as the underlying reasoning engine instead of a traditional theorem prover or SMT solver.
Our approach introduces a generic path constraint representation in terms of the original code---allowing the LLM to reason directly over the path constraint and avoiding translation into a (less expressive) formal language.
Our approach allows for a path-based decomposition of the analysis task into smaller (more tractable) subtasks, which use fewer tokens (helping {\em scale}), and are more targeted (helping {\em accuracy}).
We implemented our approach in the form of \tool---a practical LLM-based symbolic execution engine that supports multiple programming languages (\ie language agnostic, supporting \verb+C+/Python/Java) without depending on heavyweight compiler infrastructure.
Our experimental results demonstrate measurable improvements in terms of both accuracy and scale, especially in smaller models that can run on consumer-grade GPUs.

\bibliographystyle{ACM-Reference-Format}
\bibliography{main}


\begin{thebibliography}{80}

%%% the Web address.

\ifx \showCODEN    \undefined \def \showCODEN     #1{\unskip}     \fi
\ifx \showISBNx    \undefined \def \showISBNx     #1{\unskip}     \fi
\ifx \showISBNxiii \undefined \def \showISBNxiii  #1{\unskip}     \fi
\ifx \showISSN     \undefined \def \showISSN      #1{\unskip}     \fi
\ifx \showLCCN     \undefined \def \showLCCN      #1{\unskip}     \fi
\ifx \shownote     \undefined \def \shownote      #1{#1}          \fi
\ifx \showarticletitle \undefined \def \showarticletitle #1{#1}   \fi
\ifx \showURL      \undefined \def \showURL       {\relax}        \fi
% invisible to TeX
\providecommand\bibfield[2]{#2}
\providecommand\bibinfo[2]{#2}
\providecommand\natexlab[1]{#1}
\providecommand\showeprint[2][]{arXiv:#2}

\bibitem[Abdin and et~al.(2024)]%
        {Phi-4}
\bibfield{author}{\bibinfo{person}{M. Abdin} {and} \bibinfo{person}{et al.}} \bibinfo{year}{2024}\natexlab{}.
\newblock \bibinfo{title}{Phi-4 Technical Report}.
\newblock
\showeprint[arxiv]{2412.08905}~[cs.CL]
\urldef\tempurl%
\url{https://arxiv.org/abs/2412.08905}
\showURL{%
\tempurl}


\bibitem[Agrawal et~al\mbox{.}(1993)]%
        {Agrawal93}
\bibfield{author}{\bibinfo{person}{H. Agrawal}, \bibinfo{person}{R. Demillo}, {and} \bibinfo{person}{E. Spafford}.} \bibinfo{year}{1993}\natexlab{}.
\newblock \showarticletitle{Debugging with dynamic slicing and backtracking}.
\newblock \bibinfo{journal}{\emph{Softw. Pract. Exper.}} \bibinfo{volume}{23}, \bibinfo{number}{6} (\bibinfo{date}{June} \bibinfo{year}{1993}), \bibinfo{pages}{589–616}.
\newblock
\showISSN{0038-0644}
\href{https://doi.org/10.1002/spe.4380230603}{doi:\nolinkurl{10.1002/spe.4380230603}}


\bibitem[Andrès et~al\mbox{.}(2024)]%
        {Andr24}
\bibfield{author}{\bibinfo{person}{L. Andrès}, \bibinfo{person}{F. Marque}, \bibinfo{person}{A. Carcano}, \bibinfo{person}{P. Chambart}, \bibinfo{person}{J. Santos}, {and} \bibinfo{person}{J. Filliâtre}.} \bibinfo{year}{2024}\natexlab{}.
\newblock \showarticletitle{Owi: Performant Parallel Symbolic Execution Made Easy, an Application to WebAssembly}.
\newblock \bibinfo{journal}{\emph{The Art, Science, and Engineering of Programming}} \bibinfo{volume}{9}, \bibinfo{number}{1} (\bibinfo{date}{Oct.} \bibinfo{year}{2024}).
\newblock
\showISSN{2473-7321}
\href{https://doi.org/10.22152/programming-journal.org/2025/9/3}{doi:\nolinkurl{10.22152/programming-journal.org/2025/9/3}}


\bibitem[Arnold et~al\mbox{.}(2005)]%
        {Arnold05}
\bibfield{author}{\bibinfo{person}{K. Arnold}, \bibinfo{person}{J. Gosling}, {and} \bibinfo{person}{D. Holmes}.} \bibinfo{year}{2005}\natexlab{}.
\newblock \bibinfo{booktitle}{\emph{The Java programming language}}.
\newblock \bibinfo{publisher}{Addison Wesley Professional}.
\newblock


\bibitem[Asmita et~al\mbox{.}(2024)]%
        {Asmita24}
\bibfield{author}{\bibinfo{person}{Asmita}, \bibinfo{person}{Y. Oliinyk}, \bibinfo{person}{M. Scott}, \bibinfo{person}{R. Tsang}, \bibinfo{person}{C. Fang}, {and} \bibinfo{person}{H. Homayoun}.} \bibinfo{year}{2024}\natexlab{}.
\newblock \showarticletitle{Fuzzing {BusyBox}: Leveraging {LLM} and Crash Reuse for Embedded Bug Unearthing}. In \bibinfo{booktitle}{\emph{33rd USENIX Security Symposium (USENIX Security 24)}}. \bibinfo{publisher}{USENIX Association}, \bibinfo{address}{Philadelphia, PA}, \bibinfo{pages}{883--900}.
\newblock
\showISBNx{978-1-939133-44-1}
\urldef\tempurl%
\url{https://www.usenix.org/conference/usenixsecurity24/presentation/asmita}
\showURL{%
\tempurl}


\bibitem[Baldoni et~al\mbox{.}(2018)]%
        {Baldoni18}
\bibfield{author}{\bibinfo{person}{R. Baldoni}, \bibinfo{person}{E. Coppa}, \bibinfo{person}{D. D’elia}, \bibinfo{person}{C. Demetrescu}, {and} \bibinfo{person}{I. Finocchi}.} \bibinfo{year}{2018}\natexlab{}.
\newblock \showarticletitle{A Survey of Symbolic Execution Techniques}.
\newblock \bibinfo{journal}{\emph{ACM Comput. Surv.}} \bibinfo{volume}{51}, \bibinfo{number}{3}, Article \bibinfo{articleno}{50} (\bibinfo{date}{May} \bibinfo{year}{2018}), \bibinfo{numpages}{39}~pages.
\newblock
\showISSN{0360-0300}
\href{https://doi.org/10.1145/3182657}{doi:\nolinkurl{10.1145/3182657}}


\bibitem[Ball and Daniel(2015)]%
        {PyExZ3}
\bibfield{author}{\bibinfo{person}{T. Ball} {and} \bibinfo{person}{J. Daniel}.} \bibinfo{year}{2015}\natexlab{}.
\newblock \showarticletitle{Deconstructing Dynamic Symbolic Execution}.
\newblock \bibinfo{journal}{\emph{Dependable Software Systems Engineering}}  \bibinfo{volume}{40} (\bibinfo{date}{January} \bibinfo{year}{2015}).
\newblock


\bibitem[Bann(2016)]%
        {pySym}
\bibfield{author}{\bibinfo{person}{M. Bann}.} \bibinfo{year}{2016}\natexlab{}.
\newblock \bibinfo{title}{pySym: Python Symbolic Execution}.
\newblock \bibinfo{howpublished}{\url{https://github.com/bannsec/pySym}}.
\newblock


\bibitem[Brunsfeld and et~al.(2025)]%
        {tree_sitter}
\bibfield{author}{\bibinfo{person}{M. Brunsfeld} {and} \bibinfo{person}{et al.}} \bibinfo{year}{2025}\natexlab{}.
\newblock \bibinfo{booktitle}{\emph{tree-sitter/tree-sitter: v0.25.1}}.
\newblock
\href{https://doi.org/10.5281/zenodo.14788680}{doi:\nolinkurl{10.5281/zenodo.14788680}}


\bibitem[Bubeck et~al\mbox{.}(2023)]%
        {Gpt4}
\bibfield{author}{\bibinfo{person}{S. Bubeck}, \bibinfo{person}{V. Chandrasekaran}, \bibinfo{person}{R. Eldan}, \bibinfo{person}{J. Gehrke}, \bibinfo{person}{E. Horvitz}, \bibinfo{person}{E. Kamar}, \bibinfo{person}{P. Lee}, \bibinfo{person}{Y. Lee}, \bibinfo{person}{Y. Li}, \bibinfo{person}{S. Lundberg}, \bibinfo{person}{H. Nori}, \bibinfo{person}{H. Palangi}, \bibinfo{person}{M. Ribeiro}, {and} \bibinfo{person}{Y. Zhang}.} \bibinfo{year}{2023}\natexlab{}.
\newblock \bibinfo{title}{Sparks of Artificial General Intelligence: Early experiments with GPT-4}.
\newblock
\showeprint[arxiv]{2303.12712}~[cs.CL]
\urldef\tempurl%
\url{https://arxiv.org/abs/2303.12712}
\showURL{%
\tempurl}


\bibitem[Cadar et~al\mbox{.}(2008)]%
        {Cadar08}
\bibfield{author}{\bibinfo{person}{C. Cadar}, \bibinfo{person}{D. Dunbar}, {and} \bibinfo{person}{D. Engler}.} \bibinfo{year}{2008}\natexlab{}.
\newblock \showarticletitle{{KLEE}: Unassisted and Automatic Generation of {High-Coverage} Tests for Complex Systems Programs}. In \bibinfo{booktitle}{\emph{8th USENIX Symposium on Operating Systems Design and Implementation (OSDI 08)}}. \bibinfo{publisher}{USENIX Association}, \bibinfo{address}{San Diego, CA}.
\newblock
\urldef\tempurl%
\url{https://www.usenix.org/conference/osdi-08/klee-unassisted-and-automatic-generation-high-coverage-tests-complex-systems}
\showURL{%
\tempurl}


\bibitem[Cadar and Sen(2013)]%
        {Cadar13}
\bibfield{author}{\bibinfo{person}{C. Cadar} {and} \bibinfo{person}{K. Sen}.} \bibinfo{year}{2013}\natexlab{}.
\newblock \showarticletitle{Symbolic execution for software testing: three decades later}.
\newblock \bibinfo{journal}{\emph{Commun. ACM}} \bibinfo{volume}{56}, \bibinfo{number}{2} (\bibinfo{date}{Feb.} \bibinfo{year}{2013}), \bibinfo{pages}{82–90}.
\newblock
\showISSN{0001-0782}
\href{https://doi.org/10.1145/2408776.2408795}{doi:\nolinkurl{10.1145/2408776.2408795}}


\bibitem[Chen et~al\mbox{.}(2024b)]%
        {Chen24}
\bibfield{author}{\bibinfo{person}{J. Chen}, \bibinfo{person}{Z. Pan}, \bibinfo{person}{X. Hu}, \bibinfo{person}{Z. Li}, \bibinfo{person}{G. Li}, {and} \bibinfo{person}{X. Xia}.} \bibinfo{year}{2024}\natexlab{b}.
\newblock \bibinfo{title}{Reasoning Runtime Behavior of a Program with LLM: How Far Are We?}
\newblock
\showeprint[arxiv]{2403.16437}~[cs.SE]
\urldef\tempurl%
\url{https://arxiv.org/abs/2403.16437}
\showURL{%
\tempurl}


\bibitem[Chen and et~al.(2021)]%
        {Chen21}
\bibfield{author}{\bibinfo{person}{M. Chen} {and} \bibinfo{person}{et al.}} \bibinfo{year}{2021}\natexlab{}.
\newblock \bibinfo{title}{Evaluating Large Language Models Trained on Code}.
\newblock
\showeprint[arxiv]{2107.03374}~[cs.LG]
\urldef\tempurl%
\url{https://arxiv.org/abs/2107.03374}
\showURL{%
\tempurl}


\bibitem[Chen et~al\mbox{.}(2024a)]%
        {Chen24chat}
\bibfield{author}{\bibinfo{person}{Y. Chen}, \bibinfo{person}{Z. Hu}, \bibinfo{person}{C. Zhi}, \bibinfo{person}{J. Han}, \bibinfo{person}{S. Deng}, {and} \bibinfo{person}{J. Yin}.} \bibinfo{year}{2024}\natexlab{a}.
\newblock \showarticletitle{ChatUniTest: A Framework for LLM-Based Test Generation}. In \bibinfo{booktitle}{\emph{Companion Proceedings of the 32nd ACM International Conference on the Foundations of Software Engineering}} (Porto de Galinhas, Brazil) \emph{(\bibinfo{series}{FSE 2024})}. \bibinfo{publisher}{Association for Computing Machinery}, \bibinfo{address}{New York, NY, USA}, \bibinfo{pages}{572–576}.
\newblock
\showISBNx{9798400706585}
\href{https://doi.org/10.1145/3663529.3663801}{doi:\nolinkurl{10.1145/3663529.3663801}}


\bibitem[Chipounov et~al\mbox{.}(2012)]%
        {Chipounov12}
\bibfield{author}{\bibinfo{person}{V. Chipounov}, \bibinfo{person}{V. Kuznetsov}, {and} \bibinfo{person}{G. Candea}.} \bibinfo{year}{2012}\natexlab{}.
\newblock \showarticletitle{The S2E Platform: Design, Implementation, and Applications}.
\newblock \bibinfo{journal}{\emph{ACM Trans. Comput. Syst.}} \bibinfo{volume}{30}, \bibinfo{number}{1}, Article \bibinfo{articleno}{2} (\bibinfo{date}{Feb.} \bibinfo{year}{2012}), \bibinfo{numpages}{49}~pages.
\newblock
\showISSN{0734-2071}
\href{https://doi.org/10.1145/2110356.2110358}{doi:\nolinkurl{10.1145/2110356.2110358}}


\bibitem[Clarke and Emerson(1982)]%
        {Clarke81}
\bibfield{author}{\bibinfo{person}{E. Clarke} {and} \bibinfo{person}{E. Emerson}.} \bibinfo{year}{1982}\natexlab{}.
\newblock \showarticletitle{Design and synthesis of synchronization skeletons using branching time temporal logic}. In \bibinfo{booktitle}{\emph{Logics of Programs}}, \bibfield{editor}{\bibinfo{person}{Dexter Kozen}} (Ed.). \bibinfo{publisher}{Springer Berlin Heidelberg}, \bibinfo{address}{Berlin, Heidelberg}, \bibinfo{pages}{52--71}.
\newblock
\showISBNx{978-3-540-39047-3}


\bibitem[Clarke(1976)]%
        {Clarke76}
\bibfield{author}{\bibinfo{person}{L. Clarke}.} \bibinfo{year}{1976}\natexlab{}.
\newblock \showarticletitle{A System to Generate Test Data and Symbolically Execute Programs}.
\newblock \bibinfo{journal}{\emph{IEEE Transactions on Software Engineering}} \bibinfo{volume}{SE-2}, \bibinfo{number}{3} (\bibinfo{year}{1976}), \bibinfo{pages}{215--222}.
\newblock
\href{https://doi.org/10.1109/TSE.1976.233817}{doi:\nolinkurl{10.1109/TSE.1976.233817}}


\bibitem[Cousot and Cousot(1977)]%
        {Cousot77}
\bibfield{author}{\bibinfo{person}{P. Cousot} {and} \bibinfo{person}{R. Cousot}.} \bibinfo{year}{1977}\natexlab{}.
\newblock \showarticletitle{Abstract interpretation: a unified lattice model for static analysis of programs by construction or approximation of fixpoints}. In \bibinfo{booktitle}{\emph{Proceedings of the 4th ACM SIGACT-SIGPLAN Symposium on Principles of Programming Languages}} (Los Angeles, California) \emph{(\bibinfo{series}{POPL '77})}. \bibinfo{publisher}{Association for Computing Machinery}, \bibinfo{address}{New York, NY, USA}, \bibinfo{pages}{238–252}.
\newblock
\showISBNx{9781450373500}
\href{https://doi.org/10.1145/512950.512973}{doi:\nolinkurl{10.1145/512950.512973}}


\bibitem[Cousot and Cousot(1979)]%
        {Cousot79}
\bibfield{author}{\bibinfo{person}{P. Cousot} {and} \bibinfo{person}{R. Cousot}.} \bibinfo{year}{1979}\natexlab{}.
\newblock \showarticletitle{Systematic design of program analysis frameworks}. In \bibinfo{booktitle}{\emph{Proceedings of the 6th ACM SIGACT-SIGPLAN Symposium on Principles of Programming Languages}} (San Antonio, Texas) \emph{(\bibinfo{series}{POPL '79})}. \bibinfo{publisher}{Association for Computing Machinery}, \bibinfo{address}{New York, NY, USA}, \bibinfo{pages}{269–282}.
\newblock
\showISBNx{9781450373579}
\href{https://doi.org/10.1145/567752.567778}{doi:\nolinkurl{10.1145/567752.567778}}


\bibitem[de~Moura and Bjørner(2008)]%
        {DeMoura08}
\bibfield{author}{\bibinfo{person}{L. de Moura} {and} \bibinfo{person}{N. Bjørner}.} \bibinfo{year}{2008}\natexlab{}.
\newblock \showarticletitle{Z3: an efficient SMT solver}. In \bibinfo{booktitle}{\emph{2008 Tools and Algorithms for Construction and Analysis of Systems}}. \bibinfo{publisher}{Springer, Berlin, Heidelberg}, \bibinfo{pages}{337--340}.
\newblock
\urldef\tempurl%
\url{https://www.microsoft.com/en-us/research/publication/z3-an-efficient-smt-solver/}
\showURL{%
\tempurl}


\bibitem[DeepSeek-AI(2025)]%
        {DeepSeek-R1}
\bibfield{author}{\bibinfo{person}{DeepSeek-AI}.} \bibinfo{year}{2025}\natexlab{}.
\newblock \bibinfo{title}{DeepSeek-R1: Incentivizing Reasoning Capability in LLMs via Reinforcement Learning}.
\newblock
\showeprint[arxiv]{2501.12948}~[cs.CL]
\urldef\tempurl%
\url{https://arxiv.org/abs/2501.12948}
\showURL{%
\tempurl}


\bibitem[Dijkstra(1975)]%
        {Dijkstra75}
\bibfield{author}{\bibinfo{person}{E. Dijkstra}.} \bibinfo{year}{1975}\natexlab{}.
\newblock \showarticletitle{Guarded commands, nondeterminancy and formal derivation of programs}.
\newblock \bibinfo{journal}{\emph{Commun. ACM}} \bibinfo{volume}{18}, \bibinfo{number}{8} (\bibinfo{year}{1975}), \bibinfo{pages}{453--457}.
\newblock
\showISSN{0001-0782}
\href{https://doi.org/10.1145/360933.360975}{doi:\nolinkurl{10.1145/360933.360975}}


\bibitem[Distefano et~al\mbox{.}(2019)]%
        {Distefano19}
\bibfield{author}{\bibinfo{person}{D. Distefano}, \bibinfo{person}{M. F\"{a}hndrich}, \bibinfo{person}{F. Logozzo}, {and} \bibinfo{person}{P. O'Hearn}.} \bibinfo{year}{2019}\natexlab{}.
\newblock \showarticletitle{Scaling static analyses at Facebook}.
\newblock \bibinfo{journal}{\emph{Commun. ACM}} \bibinfo{volume}{62}, \bibinfo{number}{8} (\bibinfo{date}{July} \bibinfo{year}{2019}), \bibinfo{pages}{62–70}.
\newblock
\showISSN{0001-0782}
\href{https://doi.org/10.1145/3338112}{doi:\nolinkurl{10.1145/3338112}}


\bibitem[Engler et~al\mbox{.}(2001)]%
        {Dawson01}
\bibfield{author}{\bibinfo{person}{D. Engler}, \bibinfo{person}{D. Chen}, \bibinfo{person}{S. Hallem}, \bibinfo{person}{A. Chou}, {and} \bibinfo{person}{B. Chelf}.} \bibinfo{year}{2001}\natexlab{}.
\newblock \showarticletitle{Bugs as deviant behavior: a general approach to inferring errors in systems code}.
\newblock \bibinfo{journal}{\emph{SIGOPS Oper. Syst. Rev.}} \bibinfo{volume}{35}, \bibinfo{number}{5} (\bibinfo{date}{Oct.} \bibinfo{year}{2001}), \bibinfo{pages}{57–72}.
\newblock
\showISSN{0163-5980}
\href{https://doi.org/10.1145/502059.502041}{doi:\nolinkurl{10.1145/502059.502041}}


\bibitem[Fan et~al\mbox{.}(2023)]%
        {Fan23}
\bibfield{author}{\bibinfo{person}{Z. Fan}, \bibinfo{person}{X. Gao}, \bibinfo{person}{M. Mirchev}, \bibinfo{person}{A. Roychoudhury}, {and} \bibinfo{person}{S. Tan}.} \bibinfo{year}{2023}\natexlab{}.
\newblock \showarticletitle{Automated Repair of Programs from Large Language Models}. In \bibinfo{booktitle}{\emph{Proceedings of the 45th International Conference on Software Engineering}} (Melbourne, Victoria, Australia) \emph{(\bibinfo{series}{ICSE '23})}. \bibinfo{publisher}{IEEE Press}, \bibinfo{pages}{1469–1481}.
\newblock
\showISBNx{9781665457019}
\href{https://doi.org/10.1109/ICSE48619.2023.00128}{doi:\nolinkurl{10.1109/ICSE48619.2023.00128}}


\bibitem[Fang et~al\mbox{.}(2024)]%
        {Fang24}
\bibfield{author}{\bibinfo{person}{C. Fang}, \bibinfo{person}{N. Miao}, \bibinfo{person}{S. Srivastav}, \bibinfo{person}{J. Liu}, \bibinfo{person}{R. Zhang}, \bibinfo{person}{R. Fang}, \bibinfo{person}{Asmita}, \bibinfo{person}{R. Tsang}, \bibinfo{person}{N. Nazari}, \bibinfo{person}{H. Wang}, {and} \bibinfo{person}{H. Homayoun}.} \bibinfo{year}{2024}\natexlab{}.
\newblock \showarticletitle{Large language models for code analysis: do LLMs really do their job?}. In \bibinfo{booktitle}{\emph{Proceedings of the 33rd USENIX Conference on Security Symposium}} (Philadelphia, PA, USA) \emph{(\bibinfo{series}{SEC '24})}. \bibinfo{publisher}{USENIX Association}, \bibinfo{address}{USA}, Article \bibinfo{articleno}{47}, \bibinfo{numpages}{18}~pages.
\newblock
\showISBNx{978-1-939133-44-1}


\bibitem[Floyd(1967)]%
        {Floyd67}
\bibfield{author}{\bibinfo{person}{R. Floyd}.} \bibinfo{year}{1967}\natexlab{}.
\newblock \showarticletitle{Assigning Meanings to Programs}.
\newblock \bibinfo{journal}{\emph{Proceedings of Symposium on Applied Mathematics}}  \bibinfo{volume}{19} (\bibinfo{year}{1967}), \bibinfo{pages}{19--32}.
\newblock
\urldef\tempurl%
\url{http://laser.cs.umass.edu/courses/cs521-621.Spr06/papers/Floyd.pdf}
\showURL{%
\tempurl}


\bibitem[Gadelha et~al\mbox{.}(2017)]%
        {gadelha17}
\bibfield{author}{\bibinfo{person}{M. Gadelha}, \bibinfo{person}{H. Ismail}, {and} \bibinfo{person}{L. Cordeiro}.} \bibinfo{year}{2017}\natexlab{}.
\newblock \showarticletitle{Handling loops in bounded model checking of C programs via k-induction}.
\newblock \bibinfo{journal}{\emph{Int. J. Softw. Tools Technol. Transf.}} \bibinfo{volume}{19}, \bibinfo{number}{1} (\bibinfo{date}{Feb.} \bibinfo{year}{2017}), \bibinfo{pages}{97–114}.
\newblock
\showISSN{1433-2779}
\href{https://doi.org/10.1007/s10009-015-0407-9}{doi:\nolinkurl{10.1007/s10009-015-0407-9}}


\bibitem[Godefroid et~al\mbox{.}(2005)]%
        {Dart05}
\bibfield{author}{\bibinfo{person}{P. Godefroid}, \bibinfo{person}{N. Klarlund}, {and} \bibinfo{person}{K. Sen}.} \bibinfo{year}{2005}\natexlab{}.
\newblock \showarticletitle{DART: directed automated random testing}. In \bibinfo{booktitle}{\emph{Proceedings of the 2005 ACM SIGPLAN Conference on Programming Language Design and Implementation}} (Chicago, IL, USA) \emph{(\bibinfo{series}{PLDI '05})}. \bibinfo{publisher}{Association for Computing Machinery}, \bibinfo{address}{New York, NY, USA}, \bibinfo{pages}{213–223}.
\newblock
\showISBNx{1595930566}
\href{https://doi.org/10.1145/1065010.1065036}{doi:\nolinkurl{10.1145/1065010.1065036}}


\bibitem[Goodenough and Gerhart(1975)]%
        {Goodenough75}
\bibfield{author}{\bibinfo{person}{J. Goodenough} {and} \bibinfo{person}{S. Gerhart}.} \bibinfo{year}{1975}\natexlab{}.
\newblock \showarticletitle{Toward a theory of test data selection}. In \bibinfo{booktitle}{\emph{Proceedings of the International Conference on Reliable Software}} (Los Angeles, California). \bibinfo{publisher}{Association for Computing Machinery}, \bibinfo{address}{New York, NY, USA}, \bibinfo{pages}{493–510}.
\newblock
\showISBNx{9781450373852}
\href{https://doi.org/10.1145/800027.808473}{doi:\nolinkurl{10.1145/800027.808473}}


\bibitem[Graham et~al\mbox{.}(1982)]%
        {Gprof82}
\bibfield{author}{\bibinfo{person}{S. Graham}, \bibinfo{person}{P. Kessler}, {and} \bibinfo{person}{M. Mckusick}.} \bibinfo{year}{1982}\natexlab{}.
\newblock \showarticletitle{Gprof: A call graph execution profiler}. In \bibinfo{booktitle}{\emph{Proceedings of the 1982 SIGPLAN Symposium on Compiler Construction}} (Boston, Massachusetts, USA) \emph{(\bibinfo{series}{SIGPLAN '82})}. \bibinfo{publisher}{Association for Computing Machinery}, \bibinfo{address}{New York, NY, USA}, \bibinfo{pages}{120–126}.
\newblock
\showISBNx{0897910745}
\href{https://doi.org/10.1145/800230.806987}{doi:\nolinkurl{10.1145/800230.806987}}


\bibitem[Grattafiori and et~al.(2024)]%
        {Llama3}
\bibfield{author}{\bibinfo{person}{A. Grattafiori} {and} \bibinfo{person}{et al.}} \bibinfo{year}{2024}\natexlab{}.
\newblock \bibinfo{title}{The Llama 3 Herd of Models}.
\newblock
\showeprint[arxiv]{2407.21783}~[cs.AI]
\urldef\tempurl%
\url{https://arxiv.org/abs/2407.21783}
\showURL{%
\tempurl}


\bibitem[Hamlet(1977)]%
        {Hamlet77}
\bibfield{author}{\bibinfo{person}{R. Hamlet}.} \bibinfo{year}{1977}\natexlab{}.
\newblock \showarticletitle{Testing Programs with the Aid of a Compiler}.
\newblock \bibinfo{journal}{\emph{IEEE Trans. Softw. Eng.}} \bibinfo{volume}{3}, \bibinfo{number}{4} (\bibinfo{date}{July} \bibinfo{year}{1977}), \bibinfo{pages}{279–290}.
\newblock
\showISSN{0098-5589}
\href{https://doi.org/10.1109/TSE.1977.231145}{doi:\nolinkurl{10.1109/TSE.1977.231145}}


\bibitem[Hoare(1969)]%
        {Hoare69}
\bibfield{author}{\bibinfo{person}{C. Hoare}.} \bibinfo{year}{1969}\natexlab{}.
\newblock \showarticletitle{An axiomatic basis for computer programming}.
\newblock \bibinfo{journal}{\emph{Commun. ACM}} \bibinfo{volume}{12}, \bibinfo{number}{10} (\bibinfo{date}{Oct.} \bibinfo{year}{1969}), \bibinfo{pages}{576–580}.
\newblock
\showISSN{0001-0782}
\href{https://doi.org/10.1145/363235.363259}{doi:\nolinkurl{10.1145/363235.363259}}


\bibitem[Jacobs et~al\mbox{.}(2011)]%
        {Jacobs11}
\bibfield{author}{\bibinfo{person}{B. Jacobs}, \bibinfo{person}{J. Smans}, \bibinfo{person}{P. Philippaerts}, \bibinfo{person}{F. Vogels}, \bibinfo{person}{W. Penninckx}, {and} \bibinfo{person}{F. Piessens}.} \bibinfo{year}{2011}\natexlab{}.
\newblock \showarticletitle{VeriFast: A Powerful, Sound, Predictable, Fast Verifier for C and Java}. In \bibinfo{booktitle}{\emph{NASA Formal Methods}}, \bibfield{editor}{\bibinfo{person}{Mihaela Bobaru}, \bibinfo{person}{Klaus Havelund}, \bibinfo{person}{Gerard~J. Holzmann}, {and} \bibinfo{person}{Rajeev Joshi}} (Eds.). \bibinfo{publisher}{Springer Berlin Heidelberg}, \bibinfo{address}{Berlin, Heidelberg}, \bibinfo{pages}{41--55}.
\newblock
\showISBNx{978-3-642-20398-5}


\bibitem[Jain(2025)]%
        {tiktoken}
\bibfield{author}{\bibinfo{person}{S. Jain}.} \bibinfo{year}{2025}\natexlab{}.
\newblock \bibinfo{title}{tiktoken: A fast BPE tokeniser for use with OpenAI's models.}
\newblock \bibinfo{howpublished}{\url{https://github.com/openai/tiktoken}}.
\newblock


\bibitem[Kamath et~al\mbox{.}(2023)]%
        {kamath23}
\bibfield{author}{\bibinfo{person}{A. Kamath}, \bibinfo{person}{A. Senthilnathan}, \bibinfo{person}{S. Chakraborty}, \bibinfo{person}{P. Deligiannis}, \bibinfo{person}{S. Lahiri}, \bibinfo{person}{A. Lal}, \bibinfo{person}{A. Rastogi}, \bibinfo{person}{S. Roy}, {and} \bibinfo{person}{R. Sharma}.} \bibinfo{year}{2023}\natexlab{}.
\newblock \bibinfo{title}{Finding Inductive Loop Invariants using Large Language Models}.
\newblock
\showeprint[arxiv]{2311.07948}~[cs.PL]
\urldef\tempurl%
\url{https://arxiv.org/abs/2311.07948}
\showURL{%
\tempurl}


\bibitem[Kaur and Nayyar(2020)]%
        {Kaur20}
\bibfield{author}{\bibinfo{person}{A. Kaur} {and} \bibinfo{person}{R. Nayyar}.} \bibinfo{year}{2020}\natexlab{}.
\newblock \showarticletitle{A Comparative Study of Static Code Analysis tools for Vulnerability Detection in C/C++ and JAVA Source Code}.
\newblock \bibinfo{journal}{\emph{Procedia Computer Science}}  \bibinfo{volume}{171} (\bibinfo{date}{01} \bibinfo{year}{2020}), \bibinfo{pages}{2023--2029}.
\newblock
\href{https://doi.org/10.1016/j.procs.2020.04.217}{doi:\nolinkurl{10.1016/j.procs.2020.04.217}}


\bibitem[King(1976)]%
        {King76}
\bibfield{author}{\bibinfo{person}{J. King}.} \bibinfo{year}{1976}\natexlab{}.
\newblock \showarticletitle{Symbolic execution and program testing}.
\newblock \bibinfo{journal}{\emph{Commun. ACM}} \bibinfo{volume}{19}, \bibinfo{number}{7} (\bibinfo{date}{July} \bibinfo{year}{1976}), \bibinfo{pages}{385–394}.
\newblock
\showISSN{0001-0782}
\href{https://doi.org/10.1145/360248.360252}{doi:\nolinkurl{10.1145/360248.360252}}


\bibitem[Korel and Laski(1988)]%
        {Bogdan88}
\bibfield{author}{\bibinfo{person}{B. Korel} {and} \bibinfo{person}{J. Laski}.} \bibinfo{year}{1988}\natexlab{}.
\newblock \showarticletitle{Dynamic program slicing}.
\newblock \bibinfo{journal}{\emph{Inform. Process. Lett.}} \bibinfo{volume}{29}, \bibinfo{number}{3} (\bibinfo{year}{1988}), \bibinfo{pages}{155--163}.
\newblock
\showISSN{0020-0190}
\href{https://doi.org/10.1016/0020-0190(88)90054-3}{doi:\nolinkurl{10.1016/0020-0190(88)90054-3}}


\bibitem[Lattner and Adve(2004)]%
        {Chris04}
\bibfield{author}{\bibinfo{person}{C. Lattner} {and} \bibinfo{person}{V. Adve}.} \bibinfo{year}{2004}\natexlab{}.
\newblock \showarticletitle{{LLVM}: A Compilation Framework for Lifelong Program Analysis and Transformation}. \bibinfo{address}{San Jose, CA, USA}, \bibinfo{pages}{75--88}.
\newblock


\bibitem[Le~Goues et~al\mbox{.}(2012)]%
        {Genprog}
\bibfield{author}{\bibinfo{person}{C. Le~Goues}, \bibinfo{person}{T. Nguyen}, \bibinfo{person}{S. Forrest}, {and} \bibinfo{person}{W. Weimer}.} \bibinfo{year}{2012}\natexlab{}.
\newblock \showarticletitle{GenProg: A Generic Method for Automatic Software Repair}.
\newblock \bibinfo{journal}{\emph{IEEE Transactions on Software Engineering}} \bibinfo{volume}{38}, \bibinfo{number}{1} (\bibinfo{year}{2012}), \bibinfo{pages}{54--72}.
\newblock
\href{https://doi.org/10.1109/TSE.2011.104}{doi:\nolinkurl{10.1109/TSE.2011.104}}


\bibitem[Levy et~al\mbox{.}(2024)]%
        {Levy24}
\bibfield{author}{\bibinfo{person}{M. Levy}, \bibinfo{person}{A. Jacoby}, {and} \bibinfo{person}{Y. Goldberg}.} \bibinfo{year}{2024}\natexlab{}.
\newblock \showarticletitle{Same Task, More Tokens: the Impact of Input Length on the Reasoning Performance of Large Language Models}. In \bibinfo{booktitle}{\emph{Proceedings of the 62nd Annual Meeting of the Association for Computational Linguistics (Volume 1: Long Papers)}}, \bibfield{editor}{\bibinfo{person}{Lun-Wei Ku}, \bibinfo{person}{Andre Martins}, {and} \bibinfo{person}{Vivek Srikumar}} (Eds.). \bibinfo{publisher}{Association for Computational Linguistics}, \bibinfo{address}{Bangkok, Thailand}, \bibinfo{pages}{15339--15353}.
\newblock
\href{https://doi.org/10.18653/v1/2024.acl-long.818}{doi:\nolinkurl{10.18653/v1/2024.acl-long.818}}


\bibitem[Li et~al\mbox{.}(2022)]%
        {Li22}
\bibfield{author}{\bibinfo{person}{Y. Li}, \bibinfo{person}{D. Choi}, \bibinfo{person}{J. Chung}, \bibinfo{person}{N. Kushman}, \bibinfo{person}{J. Schrittwieser}, \bibinfo{person}{R. Leblond}, \bibinfo{person}{T. Eccles}, \bibinfo{person}{J. Keeling}, \bibinfo{person}{F. Gimeno}, \bibinfo{person}{A.~Dal Lago}, \bibinfo{person}{T. Hubert}, \bibinfo{person}{P. Choy}, \bibinfo{person}{C. Autume}, \bibinfo{person}{I. Babuschkin}, \bibinfo{person}{X. Chen}, \bibinfo{person}{P. Huang}, \bibinfo{person}{J. Welbl}, \bibinfo{person}{S. Gowal}, \bibinfo{person}{A. Cherepanov}, \bibinfo{person}{J. Molloy}, \bibinfo{person}{D. Mankowitz}, \bibinfo{person}{E. Robson}, \bibinfo{person}{P. Kohli}, \bibinfo{person}{N. Freitas}, \bibinfo{person}{K. Kavukcuoglu}, {and} \bibinfo{person}{O. Vinyals}.} \bibinfo{year}{2022}\natexlab{}.
\newblock \showarticletitle{Competition-level code generation with AlphaCode}.
\newblock \bibinfo{journal}{\emph{Science}} \bibinfo{volume}{378}, \bibinfo{number}{6624} (\bibinfo{date}{Dec.} \bibinfo{year}{2022}), \bibinfo{pages}{1092–1097}.
\newblock
\showISSN{1095-9203}
\href{https://doi.org/10.1126/science.abq1158}{doi:\nolinkurl{10.1126/science.abq1158}}


\bibitem[Long and Rinard(2016)]%
        {Long16}
\bibfield{author}{\bibinfo{person}{F. Long} {and} \bibinfo{person}{M. Rinard}.} \bibinfo{year}{2016}\natexlab{}.
\newblock \showarticletitle{Automatic patch generation by learning correct code}.
\newblock \bibinfo{journal}{\emph{SIGPLAN Not.}} \bibinfo{volume}{51}, \bibinfo{number}{1} (\bibinfo{date}{Jan.} \bibinfo{year}{2016}), \bibinfo{pages}{298–312}.
\newblock
\showISSN{0362-1340}
\href{https://doi.org/10.1145/2914770.2837617}{doi:\nolinkurl{10.1145/2914770.2837617}}


\bibitem[Majdoub and Charrada(2024)]%
        {Yacine24}
\bibfield{author}{\bibinfo{person}{Y. Majdoub} {and} \bibinfo{person}{E. Charrada}.} \bibinfo{year}{2024}\natexlab{}.
\newblock \showarticletitle{Debugging with Open-Source Large Language Models: An Evaluation}. In \bibinfo{booktitle}{\emph{Proceedings of the 18th ACM/IEEE International Symposium on Empirical Software Engineering and Measurement}} (Barcelona, Spain) \emph{(\bibinfo{series}{ESEM '24})}. \bibinfo{publisher}{Association for Computing Machinery}, \bibinfo{address}{New York, NY, USA}, \bibinfo{pages}{510–516}.
\newblock
\showISBNx{9798400710476}
\href{https://doi.org/10.1145/3674805.3690758}{doi:\nolinkurl{10.1145/3674805.3690758}}


\bibitem[Meng et~al\mbox{.}(2024a)]%
        {hyllfuzz}
\bibfield{author}{\bibinfo{person}{R. Meng}, \bibinfo{person}{G. Duck}, {and} \bibinfo{person}{A. Roychoudhury}.} \bibinfo{year}{2024}\natexlab{a}.
\newblock \showarticletitle{Large language model assisted hybrid fuzzing}.
\newblock \bibinfo{journal}{\emph{arXiv preprint arXiv:2412.15931}} (\bibinfo{year}{2024}).
\newblock


\bibitem[Meng et~al\mbox{.}(2024b)]%
        {Meng24}
\bibfield{author}{\bibinfo{person}{R. Meng}, \bibinfo{person}{G. Duck}, {and} \bibinfo{person}{A. Roychoudhury}.} \bibinfo{year}{2024}\natexlab{b}.
\newblock \showarticletitle{Program Environment Fuzzing}. In \bibinfo{booktitle}{\emph{Proceedings of the 2024 on ACM SIGSAC Conference on Computer and Communications Security}} (Salt Lake City, UT, USA) \emph{(\bibinfo{series}{CCS '24})}. \bibinfo{publisher}{Association for Computing Machinery}, \bibinfo{address}{New York, NY, USA}, \bibinfo{pages}{720–734}.
\newblock
\showISBNx{9798400706363}
\href{https://doi.org/10.1145/3658644.3690229}{doi:\nolinkurl{10.1145/3658644.3690229}}


\bibitem[Meng et~al\mbox{.}(2024c)]%
        {chatafl}
\bibfield{author}{\bibinfo{person}{R. Meng}, \bibinfo{person}{M. Mirchev}, \bibinfo{person}{M. B\"{o}hme}, {and} \bibinfo{person}{A. Roychoudhury}.} \bibinfo{year}{2024}\natexlab{c}.
\newblock \showarticletitle{Large Language Model guided Protocol Fuzzing}. In \bibinfo{booktitle}{\emph{Proceedings of the 31st Annual Network and Distributed System Security Symposium (NDSS)}}.
\newblock


\bibitem[Miller et~al\mbox{.}(1990)]%
        {Miller90}
\bibfield{author}{\bibinfo{person}{B. Miller}, \bibinfo{person}{L. Fredriksen}, {and} \bibinfo{person}{B. So}.} \bibinfo{year}{1990}\natexlab{}.
\newblock \showarticletitle{An empirical study of the reliability of UNIX utilities}.
\newblock \bibinfo{journal}{\emph{Commun. ACM}} \bibinfo{volume}{33}, \bibinfo{number}{12} (\bibinfo{date}{Dec.} \bibinfo{year}{1990}), \bibinfo{pages}{32–44}.
\newblock
\showISSN{0001-0782}
\href{https://doi.org/10.1145/96267.96279}{doi:\nolinkurl{10.1145/96267.96279}}


\bibitem[Nelson(2005)]%
        {nelson96}
\bibfield{author}{\bibinfo{person}{M. Nelson}.} \bibinfo{year}{2005}\natexlab{}.
\newblock \bibinfo{title}{A Survey of Reverse Engineering and Program Comprehension}.
\newblock
\showeprint[arxiv]{cs/0503068}~[cs.SE]
\urldef\tempurl%
\url{https://arxiv.org/abs/cs/0503068}
\showURL{%
\tempurl}


\bibitem[Nethercote and Seward(2007)]%
        {Valgrind07}
\bibfield{author}{\bibinfo{person}{N. Nethercote} {and} \bibinfo{person}{J. Seward}.} \bibinfo{year}{2007}\natexlab{}.
\newblock \showarticletitle{Valgrind: a framework for heavyweight dynamic binary instrumentation}.
\newblock \bibinfo{journal}{\emph{SIGPLAN Not.}} \bibinfo{volume}{42}, \bibinfo{number}{6} (\bibinfo{date}{June} \bibinfo{year}{2007}), \bibinfo{pages}{89–100}.
\newblock
\showISSN{0362-1340}
\href{https://doi.org/10.1145/1273442.1250746}{doi:\nolinkurl{10.1145/1273442.1250746}}


\bibitem[OpenAI(2024)]%
        {Openai24}
\bibfield{author}{\bibinfo{person}{OpenAI}.} \bibinfo{year}{2024}\natexlab{}.
\newblock \bibinfo{title}{GPT-4 Technical Report}.
\newblock
\showeprint[arxiv]{2303.08774}~[cs.CL]
\urldef\tempurl%
\url{https://arxiv.org/abs/2303.08774}
\showURL{%
\tempurl}


\bibitem[{OpenAI}(2024)]%
        {GPT4o-mini}
\bibfield{author}{\bibinfo{person}{{OpenAI}}.} \bibinfo{year}{2024}\natexlab{}.
\newblock \bibinfo{title}{GPT-4o mini: advancing cost-efficient intelligence}.
\newblock
\urldef\tempurl%
\url{https://openai.com/index/gpt-4o-mini-advancing-cost-efficient-intelligence/}
\showURL{%
\tempurl}


\bibitem[Pacheco and Ernst(2007)]%
        {Pacheco07}
\bibfield{author}{\bibinfo{person}{C. Pacheco} {and} \bibinfo{person}{M. Ernst}.} \bibinfo{year}{2007}\natexlab{}.
\newblock \showarticletitle{Randoop: feedback-directed random testing for Java}. In \bibinfo{booktitle}{\emph{Companion to the 22nd ACM SIGPLAN Conference on Object-Oriented Programming Systems and Applications Companion}} (Montreal, Quebec, Canada) \emph{(\bibinfo{series}{OOPSLA '07})}. \bibinfo{publisher}{Association for Computing Machinery}, \bibinfo{address}{New York, NY, USA}, \bibinfo{pages}{815–816}.
\newblock
\showISBNx{9781595938657}
\href{https://doi.org/10.1145/1297846.1297902}{doi:\nolinkurl{10.1145/1297846.1297902}}


\bibitem[P\u{a}s\u{a}reanu and Rungta(2010)]%
        {Spf10}
\bibfield{author}{\bibinfo{person}{C. P\u{a}s\u{a}reanu} {and} \bibinfo{person}{N. Rungta}.} \bibinfo{year}{2010}\natexlab{}.
\newblock \showarticletitle{Symbolic PathFinder: symbolic execution of Java bytecode}. In \bibinfo{booktitle}{\emph{Proceedings of the 25th IEEE/ACM International Conference on Automated Software Engineering}} (Antwerp, Belgium) \emph{(\bibinfo{series}{ASE '10})}. \bibinfo{publisher}{Association for Computing Machinery}, \bibinfo{address}{New York, NY, USA}, \bibinfo{pages}{179–180}.
\newblock
\showISBNx{9781450301169}
\href{https://doi.org/10.1145/1858996.1859035}{doi:\nolinkurl{10.1145/1858996.1859035}}


\bibitem[Rapps and Weyuker(1985)]%
        {Rapps85}
\bibfield{author}{\bibinfo{person}{S. Rapps} {and} \bibinfo{person}{E. Weyuker}.} \bibinfo{year}{1985}\natexlab{}.
\newblock \showarticletitle{Selecting Software Test Data Using Data Flow Information}.
\newblock \bibinfo{journal}{\emph{IEEE Transactions on Software Engineering}} \bibinfo{volume}{SE-11}, \bibinfo{number}{4} (\bibinfo{year}{1985}), \bibinfo{pages}{367--375}.
\newblock
\href{https://doi.org/10.1109/TSE.1985.232226}{doi:\nolinkurl{10.1109/TSE.1985.232226}}


\bibitem[Reynolds(2002)]%
        {Reynolds02}
\bibfield{author}{\bibinfo{person}{J. Reynolds}.} \bibinfo{year}{2002}\natexlab{}.
\newblock \showarticletitle{Separation logic: a logic for shared mutable data structures}. In \bibinfo{booktitle}{\emph{Proceedings 17th Annual IEEE Symposium on Logic in Computer Science}}. \bibinfo{pages}{55--74}.
\newblock
\href{https://doi.org/10.1109/LICS.2002.1029817}{doi:\nolinkurl{10.1109/LICS.2002.1029817}}


\bibitem[Sallou et~al\mbox{.}(2024)]%
        {Sallou24}
\bibfield{author}{\bibinfo{person}{J. Sallou}, \bibinfo{person}{T. Durieux}, {and} \bibinfo{person}{A. Panichella}.} \bibinfo{year}{2024}\natexlab{}.
\newblock \showarticletitle{Breaking the Silence: the Threats of Using LLMs in Software Engineering}. In \bibinfo{booktitle}{\emph{Proceedings of the 2024 ACM/IEEE 44th International Conference on Software Engineering: New Ideas and Emerging Results}} (Lisbon, Portugal) \emph{(\bibinfo{series}{ICSE-NIER'24})}. \bibinfo{publisher}{Association for Computing Machinery}, \bibinfo{address}{New York, NY, USA}, \bibinfo{pages}{102–106}.
\newblock
\showISBNx{9798400705007}
\href{https://doi.org/10.1145/3639476.3639764}{doi:\nolinkurl{10.1145/3639476.3639764}}


\bibitem[Schanely(2017)]%
        {CrossHair}
\bibfield{author}{\bibinfo{person}{P. Schanely}.} \bibinfo{year}{2017}\natexlab{}.
\newblock \bibinfo{title}{CrossHair: Symbolic Execution for Python}.
\newblock \bibinfo{howpublished}{\url{https://github.com/pschanely/CrossHair}}.
\newblock


\bibitem[Sen et~al\mbox{.}(2005)]%
        {Cute05}
\bibfield{author}{\bibinfo{person}{K. Sen}, \bibinfo{person}{D. Marinov}, {and} \bibinfo{person}{G. Agha}.} \bibinfo{year}{2005}\natexlab{}.
\newblock \showarticletitle{CUTE: a concolic unit testing engine for C}. In \bibinfo{booktitle}{\emph{Proceedings of the 10th European Software Engineering Conference Held Jointly with 13th ACM SIGSOFT International Symposium on Foundations of Software Engineering}} (Lisbon, Portugal) \emph{(\bibinfo{series}{ESEC/FSE-13})}. \bibinfo{publisher}{Association for Computing Machinery}, \bibinfo{address}{New York, NY, USA}, \bibinfo{pages}{263–272}.
\newblock
\showISBNx{1595930140}
\href{https://doi.org/10.1145/1081706.1081750}{doi:\nolinkurl{10.1145/1081706.1081750}}


\bibitem[Serebryany et~al\mbox{.}(2012)]%
        {Asan12}
\bibfield{author}{\bibinfo{person}{K. Serebryany}, \bibinfo{person}{D. Bruening}, \bibinfo{person}{A. Potapenko}, {and} \bibinfo{person}{D. Vyukov}.} \bibinfo{year}{2012}\natexlab{}.
\newblock \showarticletitle{AddressSanitizer: a fast address sanity checker}. In \bibinfo{booktitle}{\emph{Proceedings of the 2012 USENIX Conference on Annual Technical Conference}} (Boston, MA) \emph{(\bibinfo{series}{USENIX ATC'12})}. \bibinfo{publisher}{USENIX Association}, \bibinfo{address}{USA}, \bibinfo{pages}{28}.
\newblock


\bibitem[Shoshitaishvili et~al\mbox{.}(2016)]%
        {Angr}
\bibfield{author}{\bibinfo{person}{Y. Shoshitaishvili}, \bibinfo{person}{R. Wang}, \bibinfo{person}{C. Salls}, \bibinfo{person}{N. Stephens}, \bibinfo{person}{M. Polino}, \bibinfo{person}{A. Dutcher}, \bibinfo{person}{J. Grosen}, \bibinfo{person}{S. Feng}, \bibinfo{person}{C. Hauser}, \bibinfo{person}{C. Kruegel}, {and} \bibinfo{person}{G. Vigna}.} \bibinfo{year}{2016}\natexlab{}.
\newblock \showarticletitle{SOK: (State of) The Art of War: Offensive Techniques in Binary Analysis}. In \bibinfo{booktitle}{\emph{2016 IEEE Symposium on Security and Privacy (SP)}}. \bibinfo{pages}{138--157}.
\newblock
\href{https://doi.org/10.1109/SP.2016.17}{doi:\nolinkurl{10.1109/SP.2016.17}}


\bibitem[Si et~al\mbox{.}(2018)]%
        {Si18}
\bibfield{author}{\bibinfo{person}{X. Si}, \bibinfo{person}{H. Dai}, \bibinfo{person}{M. Raghothaman}, \bibinfo{person}{M. Naik}, {and} \bibinfo{person}{L. Song}.} \bibinfo{year}{2018}\natexlab{}.
\newblock \showarticletitle{Learning loop invariants for program verification}.
\newblock \bibinfo{journal}{\emph{Advances in Neural Information Processing Systems}}  \bibinfo{volume}{31} (\bibinfo{year}{2018}).
\newblock


\bibitem[Team(2024)]%
        {Gemma3}
\bibfield{author}{\bibinfo{person}{Gemma Team}.} \bibinfo{year}{2024}\natexlab{}.
\newblock \bibinfo{title}{Gemma: Open Models Based on Gemini Research and Technology}.
\newblock
\showeprint[arxiv]{2403.08295}~[cs.CL]
\urldef\tempurl%
\url{https://arxiv.org/abs/2403.08295}
\showURL{%
\tempurl}


\bibitem[Touvron et~al\mbox{.}(2023)]%
        {Llama23}
\bibfield{author}{\bibinfo{person}{H. Touvron}, \bibinfo{person}{T. Lavril}, \bibinfo{person}{G. Izacard}, \bibinfo{person}{X. Martinet}, \bibinfo{person}{M. Lachaux}, \bibinfo{person}{T. Lacroix}, \bibinfo{person}{B. Rozière}, \bibinfo{person}{N. Goyal}, \bibinfo{person}{E. Hambro}, \bibinfo{person}{F. Azhar}, \bibinfo{person}{A. Rodriguez}, \bibinfo{person}{A. Joulin}, \bibinfo{person}{E. Grave}, {and} \bibinfo{person}{G. Lample}.} \bibinfo{year}{2023}\natexlab{}.
\newblock \bibinfo{title}{LLaMA: Open and Efficient Foundation Language Models}.
\newblock
\showeprint[arxiv]{2302.13971}~[cs.CL]
\urldef\tempurl%
\url{https://arxiv.org/abs/2302.13971}
\showURL{%
\tempurl}


\bibitem[van Rossum(1995)]%
        {Python}
\bibfield{author}{\bibinfo{person}{G. van Rossum}.} \bibinfo{year}{1995}\natexlab{}.
\newblock \bibinfo{booktitle}{\emph{Python tutorial}}.
\newblock \bibinfo{type}{{T}echnical {R}eport} CS-R9526. \bibinfo{institution}{Centrum voor Wiskunde en Informatica (CWI)}, \bibinfo{address}{Amsterdam}.
\newblock


\bibitem[Visser et~al\mbox{.}(2004)]%
        {Visser04}
\bibfield{author}{\bibinfo{person}{W. Visser}, \bibinfo{person}{C. Păsăreanu}, {and} \bibinfo{person}{S. Khurshid}.} \bibinfo{year}{2004}\natexlab{}.
\newblock \showarticletitle{Test input generation with java PathFinder}. In \bibinfo{booktitle}{\emph{Proceedings of the 2004 ACM SIGSOFT International Symposium on Software Testing and Analysis}} (Boston, Massachusetts, USA) \emph{(\bibinfo{series}{ISSTA '04})}. \bibinfo{publisher}{Association for Computing Machinery}, \bibinfo{address}{New York, NY, USA}, \bibinfo{pages}{97–107}.
\newblock
\showISBNx{1581138202}
\href{https://doi.org/10.1145/1007512.1007526}{doi:\nolinkurl{10.1145/1007512.1007526}}


\bibitem[Wang et~al\mbox{.}(2024)]%
        {wang24}
\bibfield{author}{\bibinfo{person}{W. Wang}, \bibinfo{person}{K. Liu}, \bibinfo{person}{A. Chen}, \bibinfo{person}{G. Li}, \bibinfo{person}{Z. Jin}, \bibinfo{person}{G. Huang}, {and} \bibinfo{person}{L. Ma}.} \bibinfo{year}{2024}\natexlab{}.
\newblock \bibinfo{title}{Python Symbolic Execution with LLM-powered Code Generation}.
\newblock
\showeprint[arxiv]{2409.09271}~[cs.SE]
\urldef\tempurl%
\url{https://arxiv.org/abs/2409.09271}
\showURL{%
\tempurl}


\bibitem[Weiser(1981)]%
        {Weiser81}
\bibfield{author}{\bibinfo{person}{M. Weiser}.} \bibinfo{year}{1981}\natexlab{}.
\newblock \showarticletitle{Program slicing}. In \bibinfo{booktitle}{\emph{Proceedings of the 5th International Conference on Software Engineering}} (San Diego, California, USA) \emph{(\bibinfo{series}{ICSE '81})}. \bibinfo{publisher}{IEEE Press}, \bibinfo{pages}{439–449}.
\newblock
\showISBNx{0897911466}


\bibitem[Weyuker(1983)]%
        {Weyuker83}
\bibfield{author}{\bibinfo{person}{E. Weyuker}.} \bibinfo{year}{1983}\natexlab{}.
\newblock \showarticletitle{Assessing Test Data Adequacy through Program Inference}.
\newblock \bibinfo{journal}{\emph{ACM Trans. Program. Lang. Syst.}} \bibinfo{volume}{5}, \bibinfo{number}{4} (\bibinfo{date}{Oct.} \bibinfo{year}{1983}), \bibinfo{pages}{641–655}.
\newblock
\showISSN{0164-0925}
\href{https://doi.org/10.1145/69575.357231}{doi:\nolinkurl{10.1145/69575.357231}}


\bibitem[Xi et~al\mbox{.}(2025)]%
        {Xi25}
\bibfield{author}{\bibinfo{person}{Z. Xi}, \bibinfo{person}{W. Chen}, \bibinfo{person}{X. Guo}, \bibinfo{person}{W. He}, \bibinfo{person}{Y. Ding}, \bibinfo{person}{B. Hong}, \bibinfo{person}{M. Zhang}, \bibinfo{person}{J. Wang}, \bibinfo{person}{S. Jin}, \bibinfo{person}{E. Zhou}, \bibinfo{person}{R. Zheng}, \bibinfo{person}{X. Fan}, \bibinfo{person}{X. Wang}, \bibinfo{person}{L. Xiong}, \bibinfo{person}{Y. Zhou}, \bibinfo{person}{W. Wang}, \bibinfo{person}{C. Jiang}, \bibinfo{person}{Y. Zou}, \bibinfo{person}{X. Liu}, \bibinfo{person}{Z. Yin}, \bibinfo{person}{S. Dou}, \bibinfo{person}{R. Weng}, \bibinfo{person}{W. Cheng}, \bibinfo{person}{Q. Zhang}, \bibinfo{person}{W. Qin}, \bibinfo{person}{Y. Zheng}, \bibinfo{person}{X. Qiu}, \bibinfo{person}{X. Huang}, {and} \bibinfo{person}{T. Gui}.} \bibinfo{year}{2025}\natexlab{}.
\newblock \showarticletitle{The rise and potential of large language model based agents: a survey}.
\newblock \bibinfo{journal}{\emph{Science China Information Sciences}} \bibinfo{volume}{68}, \bibinfo{number}{2} (\bibinfo{date}{17 Jan} \bibinfo{year}{2025}), \bibinfo{pages}{121101}.
\newblock
\showISSN{1869-1919}
\href{https://doi.org/10.1007/s11432-024-4222-0}{doi:\nolinkurl{10.1007/s11432-024-4222-0}}


\bibitem[Xia et~al\mbox{.}(2023)]%
        {Xia23}
\bibfield{author}{\bibinfo{person}{C.~Steven Xia}, \bibinfo{person}{Y. Wei}, {and} \bibinfo{person}{L. Zhang}.} \bibinfo{year}{2023}\natexlab{}.
\newblock \showarticletitle{Automated Program Repair in the Era of Large Pre-trained Language Models}. In \bibinfo{booktitle}{\emph{2023 IEEE/ACM 45th International Conference on Software Engineering (ICSE)}}. \bibinfo{pages}{1482--1494}.
\newblock
\href{https://doi.org/10.1109/ICSE48619.2023.00129}{doi:\nolinkurl{10.1109/ICSE48619.2023.00129}}


\bibitem[Yang and et~al.(2025)]%
        {Qwen3}
\bibfield{author}{\bibinfo{person}{A. Yang} {and} \bibinfo{person}{et al.}} \bibinfo{year}{2025}\natexlab{}.
\newblock \bibinfo{title}{Qwen3 Technical Report}.
\newblock
\showeprint[arxiv]{2505.09388}~[cs.CL]
\urldef\tempurl%
\url{https://arxiv.org/abs/2505.09388}
\showURL{%
\tempurl}


\bibitem[Zeller(2005)]%
        {Zeller05}
\bibfield{author}{\bibinfo{person}{A. Zeller}.} \bibinfo{year}{2005}\natexlab{}.
\newblock \bibinfo{booktitle}{\emph{Why Programs Fail: A Guide to Systematic Debugging}}.
\newblock \bibinfo{publisher}{Morgan Kaufmann Publishers Inc.}, \bibinfo{address}{San Francisco, CA, USA}.
\newblock
\showISBNx{1558608664}


\bibitem[Zeller et~al\mbox{.}(2024)]%
        {fuzzingbook2024}
\bibfield{author}{\bibinfo{person}{A. Zeller}, \bibinfo{person}{R. Gopinath}, \bibinfo{person}{M. B{\"o}hme}, \bibinfo{person}{G. Fraser}, {and} \bibinfo{person}{C. Holler}.} \bibinfo{year}{2024}\natexlab{}.
\newblock \bibinfo{booktitle}{\emph{The Fuzzing Book}}.
\newblock \bibinfo{publisher}{CISPA Helmholtz Center for Information Security}.
\newblock
\urldef\tempurl%
\url{https://www.fuzzingbook.org/}
\showURL{%
\tempurl}
\newblock
\shownote{Retrieved 2024-07-01 16:50:18+02:00}.


\bibitem[Zhou et~al\mbox{.}(2024a)]%
        {Lo24detect}
\bibfield{author}{\bibinfo{person}{X. Zhou}, \bibinfo{person}{S. Cao}, \bibinfo{person}{X. Sun}, {and} \bibinfo{person}{D. Lo}.} \bibinfo{year}{2024}\natexlab{a}.
\newblock \showarticletitle{Large Language Model for Vulnerability Detection and Repair: Literature Review and the Road Ahead}.
\newblock \bibinfo{journal}{\emph{ACM Trans. Softw. Eng. Methodol.}} (\bibinfo{date}{Dec.} \bibinfo{year}{2024}).
\newblock
\showISSN{1049-331X}
\href{https://doi.org/10.1145/3708522}{doi:\nolinkurl{10.1145/3708522}}
\newblock
\shownote{Just Accepted}.


\bibitem[Zhou et~al\mbox{.}(2024b)]%
        {Zhou24}
\bibfield{author}{\bibinfo{person}{X. Zhou}, \bibinfo{person}{T. Zhang}, {and} \bibinfo{person}{D. Lo}.} \bibinfo{year}{2024}\natexlab{b}.
\newblock \showarticletitle{Large Language Model for Vulnerability Detection: Emerging Results and Future Directions}. In \bibinfo{booktitle}{\emph{Proceedings of the 2024 ACM/IEEE 44th International Conference on Software Engineering: New Ideas and Emerging Results}} (Lisbon, Portugal) \emph{(\bibinfo{series}{ICSE-NIER'24})}. \bibinfo{publisher}{Association for Computing Machinery}, \bibinfo{address}{New York, NY, USA}, \bibinfo{pages}{47–51}.
\newblock
\showISBNx{9798400705007}
\href{https://doi.org/10.1145/3639476.3639762}{doi:\nolinkurl{10.1145/3639476.3639762}}


\bibitem[Zhu et~al\mbox{.}(1997)]%
        {Zhu97}
\bibfield{author}{\bibinfo{person}{H. Zhu}, \bibinfo{person}{P. Hall}, {and} \bibinfo{person}{J. May}.} \bibinfo{year}{1997}\natexlab{}.
\newblock \showarticletitle{Software unit test coverage and adequacy}.
\newblock \bibinfo{journal}{\emph{ACM Comput. Surv.}} \bibinfo{volume}{29}, \bibinfo{number}{4} (\bibinfo{date}{Dec.} \bibinfo{year}{1997}), \bibinfo{pages}{366–427}.
\newblock
\showISSN{0360-0300}
\href{https://doi.org/10.1145/267580.267590}{doi:\nolinkurl{10.1145/267580.267590}}


\end{thebibliography}

\end{document}